\documentclass{ws-mplb}
\usepackage{epsfig}

\def\avg#1{\langle#1\rangle}
\def\Re {\mbox{Re}}
\def\Im {\mbox{Im}}
\def\be{\begin{equation}}       \def\ee{\end{equation}}
\def\bea{\begin{eqnarray}}      \def\eea{\end{eqnarray}}

\def\nn{\nonumber}

\begin{document}

\markboth{Congjun Wu}{Hidden symmetry and quantum phases in spin-3/2
cold atomic systems}

\catchline{}{}{}{}{}
%
%%%%%%%%%%%%%%%%%%%%%%%%%%%%%%%%%%%%%%%%%%%%%%%%%%%%%%%%%%%%%%%%%%%%

\title{Hidden symmetry and quantum phases in spin-3/2
cold atomic systems
%\footnote{For the title, try not to 
%use more than 3 lines. Typeset the title in 10 pt 
%Times roman, uppercase and boldface.}
}

\author{\footnotesize Congjun Wu
%\footnote{Typeset names in 
%10~pt Times roman, uppercase. Use the footnote to indicate 
%the present or permanent address of the author.}
}

\address{ Kavli Institute for Theoretical Physics, University 
of California, \\  Santa  Barbara, CA 93106, USA. \\
wucj@kitp.ucsb.edu}

\maketitle

\begin{history}
\received{(Day Month Year)}
\revised{(Day Month Year)}
\end{history}

\begin{abstract}
Optical traps and lattices provide a new opportunity to study
strongly correlated high spin systems with cold atoms.
In this article, we review the recent progress on the hidden symmetry
properties in the simplest high spin fermionic systems with hyperfine
spin $F=3/2$, which may be realized with atoms of 
$^{132}$Cs, $^9$Be, $^{135}$Ba, $^{137}$Ba, and $^{201}$Hg.
A {\it generic} $SO(5)$ or isomorphically, $Sp(4)$) symmetry is proved
in such systems with the $s$-wave scattering interactions in optical traps,
or with the on-site Hubbard interactions in optical lattices. 
Various important features from this high symmetry are studied in
the Fermi liquid theory, the mean field phase diagram, 
and the sign problem in quantum Monte-Carlo simulations.
In the $s$-wave quintet Cooper pairing phase, the half-quantum vortex 
exhibits the global analogue of the Alice string and non-Abelian 
Cheshire charge properties in gauge theories. 
The existence of the quartetting phase, a four-fermion counterpart of the
Cooper pairing phase, and its competition with other orders
are studied in one dimensional spin-$3/2$ systems.
We also show that counter-intuitively quantum fluctuations in spin-3/2 magnetic
systems are even stronger than those in spin-1/2 systems.
\end{abstract}


\section{Introduction}
\label{sect:intro}
The past decade has witnessed the great progress in the cold atomic physics.
The first generation of Bose-Einstein condensation (BEC) in alkali atoms 
was realized in magnetic traps.
In such systems,  spins of atoms are fully polarized by the Zeemann field
\cite{anderson1995,davis1995}, thus atoms are effectively of single component.
A few years later, important achievement was made to release the spin
degrees of freedom by using optical methods to trap atoms, such as
optical traps and lattices 
\cite{myatt1997,stamperkurn1998,stenger1998}.
An additional advantage of optical lattices is 
the excellent  controllability  of the interaction strength 
from the weak coupling to strong coupling regimes. 
For example, the superfluid to Mott insulator transition of 
bosons  has been experimentally observed \cite{greiner2002a}. 
All these progresses provide a controllable way to study 
high spin physics, in particular, the strongly correlated high spin
physics. 

The high spin physics with cold atoms contains novel features which
does not appear in solid state systems.
In many transition metal oxides, several electrons are combined by 
Hund's rule to form an onsite composite high spin object with $S>\frac{1}{2}$.
However, the intersite coupling is still dominated by the exchange process 
of a single pair of electrons,
thus the leading order spin exchange is the bilinear Heisenberg term.
Due to the $1/S$ effect, quantum fluctuations are typically weak.
In other words, the solid state high spin  systems are
more classical than  spin-$\frac{1}{2}$ systems.
In contrast, such restriction does not happen in high spin systems 
with cold atoms  because the building block of such systems, each atom,
carries a high hyperfine spin $F$ which includes both electron 
and nuclear spins.
This feature renders the possible high hidden symmetries and
strong quantum fluctuations.

The high spin bosonic systems exhibit much richer phase diagram 
than the spinless boson.
In spin-1 systems of $^{23}$Na and $^{87}$Rb atoms,
polar and ferromagnetic spinor condensations and related
collective modes were studied by Ho \cite{ho1998} 
and Ohmi {\it et al} \cite{ohmi1998}.
The interplay between the $U(1)$ charge
and $SU(2)$ spin degrees of freedom in the spinor condensate
is characterized by a $Z_2$ gauge symmetry.
The topological aspect of this hidden $Z_2$ gauge symmetry
was investigated extensively  by Zhou
\cite{zhou2001,zhou2003} and Demler {\it et al} \cite{demler2002}.
Furthermore, the spin singlet state and spin nematic state
were also studied in optical lattices
by Zhou {\it et al.} \cite{zhou2003a} and Imambekov
{\it et al} \cite{imambekov2003}.
Recently, the spin-$2$ hyperfine state of $^{87}$Rb atom 
\cite{widera2006,zhou2006} and spin-$3$ 
atom of $^{52}$Cr \cite{Griesmaier2005,Stuhler2005,diener2006,santos2005} 
have attracted a lot of attention.
Various spinor condensates and spin ordered Mott insulating states
have been classified.
Remarkably, a biaxial spin-nematic state can be stabilized 
in $^{52}$Cr systems which
can support interesting topological defects of non-Abelian vortices.

On the other hand, relatively less works have been done for
high spin fermions.
However, such systems indeed have interesting
properties, and thus deserve attention.
Pioneering works by Yip {\it et al.} \cite{yip1999}
and Ho {\it et al.} \cite{yip1999} showed that high spin fermionic
systems exhibit richer structures of collective modes in the 
Fermi liquid theory, and more diversities in Cooper pairing patterns.
More recently, a large progress has been made by Wu {\it et al.} \cite{wu2003}
on the symmetry properties in spin-$3/2$ systems,
which may be realized with atoms such as
$^{132}$Cs, $^9$Be, $^{135}$Ba, $^{137}$Ba, $^{201}$Hg.
Such systems are very special in that in additional
to the obvious spin $SU(2)$ symmetry, they possess
a hidden and {\it generic} $SO(5)$ or isomorphically, $Sp(4)$ symmetry.
In this paper, we will review this progress.

Before we move on, let us briefly review some group theory knowledge.
The $SO(5)$ group describes the rotation in the 5-dimensional space
spanned by five real axes $n_{1}\sim n_5$ which form
the vector representation of $SO(5)$.
$SO(5)$ group has ten generators $L_{ab}~ (1\le a<b\le 5)$, each of which
is responsible for the rotation in the $ab$ plane.
These generators form $SO(5)$'s 10 dimensional adjoint representation.
$SO(5)$'s fundamental spinor representation is 4-dimensional.
Rigorously speaking, its spinor representation is faithful for
$SO(5)$'s covering group $Sp(4)$.
The relation between $SO(5)$ and $Sp(4)$ is similar to that between
$SO(3)$ and $SU(2)$.
For simplicity, we will not distinguish the difference between $SO(5)$
and $Sp(4)$ below.

The origin of the $SO(5)$ symmetry in spin-$3/2$ systems 
can be explained as follows:
The kinetic energy term shows an obvious $SU(4)$ symmetry because of
the equivalence among the four spin components $\psi_\alpha(\alpha=\pm 
\frac{3}{2},\pm\frac{1}{2})$.
However, this $SU(4)$ symmetry generally speaking is broken explicitly 
down to $SO(5)$ by interactions.
Pauli's exclusion principle requires that in the $s$-wave scattering 
channel, only interactions $g_0$ in the total spin singlet channel 
($S_T=0$) and $g_2$ in the quintet channel $(S_T=2)$ are allowed, 
while those in the channels of $S_T=1,3$ are forbidden.
Remarkably, the singlet and quintet channels automatically form the 
identity and 5-vector representations for an $SO(5)$ group 
(see Sect. \ref{sect:so5} for detail),
thus making the system $SO(5)$ invariant without fine tuning.
The same reasoning also applies to the existence of
the $SO(5)$ symmetry in the lattice Hubbard model.
The validity of this $SO(5)$ symmetry is regardless of
dimensionality, lattice geometry and impurity potentials.
Basically, it plays the same role of the
$SU(2)$ symmetry in the spin-$\frac{1}{2}$ systems.
This $SO(5)$ symmetry purely lies in the particle-hole
channel as an extension of the spin $SU(2)$ algebra.
It is qualitatively different from the $SO(5)$ theory in 
the high T$_c$ superconductivity \cite{zhang1997}
which involves both particle-hole and particle-particle channels.

To our knowledge, such a high symmetry without fine tuning is rare
in condensed matter and cold atomic systems, and thus it 
is worthwhile for  further exploration.
Below we outline several important consequences from the $SO(5)$
symmetry and other interesting properties in spin-$3/2$
systems.
The $SO(5)$ symmetry brings hidden degeneracy in the collective
modes in the Fermi liquid theory \cite{wu2003}.
This symmetry greatly facilitates the understanding of
the mean field phase diagram in the lattice system \cite{wu2003}.
Furthermore, due to the time-reversal properties of the $SO(5)$
algebra, the sign problem of the quantum Monte-Carlo
algorithm is  proved to be absent in a large part of the
phase diagram \cite{wu2003,wu2005}. 

Spin-$3/2$ systems can support the quintet Cooper
pairing phase i.e., Cooper pair with total spin $S_{tot}=2$.
This high spin superfluid state possesses the topological defect
of the half-quantum vortex  often called the Alice string.
An interesting property of the half quantum vortex loop or pair
is that they can carry spin quantum number as a
global analogy of the Cheshire charge in the gauge theories.
Interestingly, in the quintet pairing state,
when a quasiparticle carrying spin penetrates the
Cheshire charged half-quantum vortex loop, quantum entanglement
is generated between them.
The Alice string and non-Abelian Cheshire charge behavior in spin-$3/2$
systems were studied by Wu {\it et al} \cite{wu2005b}.

Fermionic systems with multiple components can support 
multiple-particle clustering instabilities, which means more 
than two fermions come together to form bound states
\cite{schlottmann1994,stepaneko1999,kamei2005,wu2005a,lee2006,rapp2006,honerkamp2003}.
For example, baryons are bound states of three-quarks,
$\alpha$-particles are bound states of two protons and
two neutrons, and bi-excitons are  bound state of two electrons
and two holes.
Spin-$3/2$ systems can exhibit quartetting order
as a four-fermion counterpart of the Cooper pairing.
Taking into account the arrival of the fermion pairing superfluidity
by Feshbach resonances 
\cite{regal2004,Zwierlein2004,Kinast2004,Bartenstein2004}, 
it is natural to expect the
quartetting order as a possible research focus in the near future.
The existence of the quartetting order and its competition
with the singlet pairing order were investigated 
by Wu \cite{wu2005a} in 1D spin-$3/2$ systems.
The generalization of the quartetting order to even higher
spin systems at 1D was investigated by 
Lecheminant {\it et al} \cite{lecheminant2005}.

Spin-$3/2$ magnetic systems are  characterized by strong quantum fluctuations. 
This is a little bit counter-intuitive because 
one would expect weak quantum fluctuations due to the high spin. 
However, because of the high symmetry of $SO(5)$, quantum fluctuations
are actually even stronger than those in spin $\frac{1}{2}$ systems.
For example, spin-$3/2$ systems at quarter filling
(one particle per site) can exhibit the $SU(4)$ plaquette order
which is a four-site counterpart of the dimer state in 
spin-$\frac{1}{2}$ systems.
An $SU(4)$ Majumdar-Ghosh model was constructed by Chen {\it et al.}
\cite{chen2005} in the spin-$3/2$ two-leg ladder 
systems.
Their ground state is exactly solvable and exhibits such plaquette order.

On the experimental side, in addition to alkali atoms, 
considerable progress has been made in trapping 
and cooling the divalent alkaline-earth atoms
\cite{machholm2001,maruyama2003,takasu2003,takasu2003a}.
Two candidate spin-$3/2$ atoms are alkaline-earth atoms of
$^{135}$Ba and $^{137}$Ba.
Their resonances of $6s^2\rightarrow 6s^16p^1$ are at 553.7 nm 
\cite{he1991}, thus making them as promising candidates 
for laser cooling and further experimental investigation.
Their scattering lengths are not available now, but that of $^{138}$Ba (spin 0)
was estimated as $-41 a_B$ \cite{machholm2001}.
Because the $6s$ shell of Ba is fully-filled,
both the $a_0$, $a_2$ of $^{135}$Ba and $^{137}$Ba
should have the similar value.
Considering the rapid development in this field,
we expect that more spin-3/2 systems can be realized
experimentally.  

In the rest of this paper, we will review the progress in 
spin-$3/2$ system,
and also present some new results unpublished before.
In Sect. \ref{sect:so5}, we introduce the proof of the exact $SO(5)$
symmetry.
In Sect. \ref{sect:effect}, we present the effect of the $SO(5)$ symmetry
to the spin-$3/2$ Fermi liquid theory, 
the mean field phase diagram of the lattice Hubbard model, and
the sign problem of quantum Monte Carlo sign problem.
In Sect. \ref{sect:quintet}, we review the quintet Cooper pairing and
the topological defect of the half-quantum vortex.
In Sect. \ref{sect:quartet} we discuss the quartetting instability 
and its competition with pairing instability in 1D systems.
In Sect. \ref{sect:magnetism}, we study the spin-$3/2$ magnetism.
We summarize the paper in Sect. \ref{sect:summary}.

Due to limitation of space, we will not cover several
important works on the spin-$3/2$ fermions.
For example, Hattori \cite{Hattori2005} studied the Kondo
problem finding an interesting non-Fermi liquid fixed point.
Recently, the 1D spin-$3/2$ model with arbitrary interactions
was solved using Bethe-ansatz  by Controzzi {\it et. al} 
\cite{Controzzi2006}.

%**********************************************************************
\section{The hidden $SO(5) (Sp(4))$ symmetry}
\label{sect:so5}
In this section, we will review the hidden symmetry properties of
spin-$3/2$ systems \cite{wu2003,wu2005}.
Let us first look at a familiar example of hidden
symmetry: the hydrogen atom.
The hydrogen atom has  an obvious $SO(3)$ symmetry which can not
explain the large energy level  degeneracy pattern of $n^2$.
Actually, this degeneracy is not accidental.
It arises from the $1/r$ Coulomb potential which
gives rise to a hidden conserved quantity, the 
Runge-Lentz (R-L) vector.
This vector describes the orientation and  eccentricity
of the classical elliptic orbits.
The R-L vector and orbital angular momentum together form a closed 
$SO(4)$ algebra which is responsible for the degeneracy pattern of $n^2$.
Now we can ask: what are the hidden conserved quantities
in spin-3/2 systems? We will answer this question below.

\subsection{Model Hamiltonians}
We start with the generic form of the spin-$3/2$ Hamiltonian
of the continuum model \cite{ho1999,yip1999}.
The only assumptions we will make  are the spin $SU(2)$
symmetry and the $s$-wave scattering interactions.
Required by Paul's exclusion principle, the spin channel
wave function of two fermions has to be antisymmetric in the 
$s$-partial wave channel.
As a result, only two independent interaction channels exist
with the total spin $F=0,2$.
Taking the above facts into account, the Hamiltonian reads
\bea
H&=& \int d^d{\vec r} ~ \Big\{ \sum_{\alpha=\pm 3/2, \pm 1/2}
 \psi^\dagger_\alpha({\vec r})
\big (-\frac{\hbar^2}{2m}\nabla^2-\mu\big) \psi_\alpha({\vec r}) 
+ g_0 P_{0,0}^\dagger({\vec r}) P_{0,0}({\vec r}) \nonumber \\
&+&
g_2 \sum_{m=\pm2,\pm1,0} P_{2,m}^\dagger({\vec r})
P_{2,m}({\vec r})\Big \},
\label{eq:ctnHam}
\eea
with $d$ the space dimension,  $\mu$ the chemical potential.
$P^\dagger_{0,0}, P^\dagger_{2,m}$ are the singlet ($S_T=0$) and quintet
($S_T=2$) pairing operators  defined through the Clebsh-Gordan
coefficient for two indistinguishable particles as 
$P_{F,m}^\dagger(\vec r)=\sum_{\alpha\beta} \avg{\frac{3}{2}
\frac{3}{2};F,m|\frac{3}{2}\frac{3}{2}\alpha\beta}
\psi^\dagger_\alpha(\vec r) \psi^\dagger_\beta(\vec r), 
$
where $F=0,2$ and $m=-F,-F+1, ...,  F$.
The feature that only two interaction channels exist
is important for the existence of the hidden
$SO(5)$ symmetry.

If spin-$3/2$ atoms are loaded into optical lattices, 
we need to construct the lattice Hamiltonian.
We assume $V_0$ the potential depth , $k=\pi/l_0$ the light wavevector,
and $l_0$ the lattice constant. The hopping integral $t$ 
decreases exponentially with increasing $V_0$.
Within the harmonic approximation, 
$U/\Delta E\approx (\pi^2 / 2) (a_s/l_0)
(V_0/E_r)^{1/4}$,
with $U$  the repulsion of two fermions on one
site, $\Delta E$ the gap between the lowest and first excited
single particle state in one site, $a_s$ the $s$-wave scattering
length in the corresponding channel, and $E_r= \hbar^2 k^2 /2M$  the
recoil  energy. 
In the absence of Feshbach resonances, the
typical estimation reads  $a_s\sim 100 a_B$ ( $a_B$ the Bohr radius),
$l_0\sim 5000$A, and $(V_0/E_r)^{1/4} \approx 1\sim 2$.
Thus we arrive at $U/\Delta E< 0.1$, and the system can be
approximated by the one-band Hubbard model
\bea
H&=&-t\sum_{\langle ij\rangle ,\sigma} \big \{
c^\dagger_{i\sigma}
 c_{j\sigma} +h.c.\big \}-\mu \sum_{i\sigma} c^\dagger_{i\sigma} c_{i\sigma}
+U_0 \sum_i P_{0,0}^\dagger(i) P_{0,0}(i)
\nonumber \\
&+&U_2 \sum_{i,m=\pm2,\pm1,0}
P_{2,m}^\dagger(i) P_{2,m}(i). 
\label{eq:latHam} 
\eea 
for the particle density $n\leq 4$. 
At half-filling in a bipartite lattice, $\mu$ is given by
$\mu_0=(U_0+5U_2)/4$ to ensure the particle-hole ($p$-$h$) symmetry.
The lattice fermion operators and their
continuum counterparts are related by $\psi_\alpha ({\vec r})=
c_\alpha (i)/(l_0)^{d/2}$. 

\begin{figure}
\centering\epsfig{file=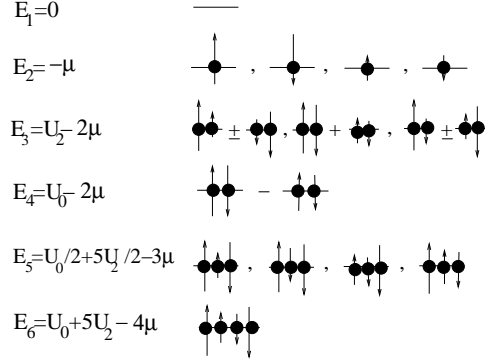,clip=1,width=0.5\linewidth}
\caption{Energy level diagram for a single site problem.
The longer and shorter arrows denote the spin components of
$S_z=\pm \frac{3}{2}$ and $S_z=\pm \frac{1}{2}$ respectively.
The sets of E$_{1,4,6}$ (spin singlet) are also $SO(5)$ singlet;
E$_{2,5}$(spin quartet) are $SO(5)$ spinors;
E$_3$ (spin quintet) is an $SO(5)$ vector.}
\label{fig:egylvl} 
\end{figure} 

The best way to understand the $SO(5)$ symmetry without going through
the mathematical details is to look at the energy level diagram
for the  single site problem  as depicted in
Fig. \ref{fig:egylvl}.
It contains 2$^4$=16 states which can be classified into three sets of
spin singlets ($E_{1,4,6}$), two sets of spin quartets ($E_{2,5}$),
and one set of spin quintet ($E_3$).
Interestingly, this energy level degeneracy pattern matches $SO(5)$
representations perfectly: the spin singlets $E_{1,4,6}$, quartets $E_{2,5}$,
and quintet $E_3$ can be considered as $SO(5)$
singlets, spinors, and vector as well.
Thus the single site spectrum is $SO(5)$ symmetric without any fine tuning.
If we further tune $U_0=U_2=U$, then $E_{3,4}$ become 6-fold degenerate
making the system $SU(4)$ invariant.
In this case, interactions in Eq. \ref{eq:latHam} reduce to
the density-density interaction as $H_{int}=\frac{U}{2} n(n-1)$. 
The $SU(4)$ symmetry simply means that the four spin-components are
equivalent to each other.

\subsection{The particle-hole channel SO(5) algebra}
In order to prove the $SO(5)$ symmetry in the many-body Hamiltonians of
Eq. \ref{eq:ctnHam} and \ref{eq:latHam}, 
we need to construct the $SO(5)$ algebra. 
The four spin components $\psi_\alpha(\vec r) (\alpha=\pm\frac{3}{2},
\pm\frac{1}{2})$  render $4^2=16$ $p$-$h$ channel bi-linears.
As a result, the charge operator and three spin $F_{x,y,z}$ 
operators are not a complete set for  describing the total degrees of freedom.
High order tensors are needed in terms of the tensor product of $s=3/2$ 
spin matrices $F_i$:
\bea \label{s32_algebra}
&\mbox{rank-0:}& I, \nonumber \\
&\mbox{rank-1:} & F^i, \ \ i=1,2,3, \nonumber \\
&\mbox{rank-2:} & \xi^a_{ij} F_i F_j, \ \ a=1,..,5, \ \ \xi^a_{ij}=\xi^a_{ji}, \ \ \xi^a_{ii}=0, \nonumber \\
&\mbox{rank-3:} & \xi^L_{ijk} F_i F_j F_k, \ \ L=1,..,7, \ \ \xi^L_{ijk}=\xi^L_{jik}, \ \ \xi^L_{iik}=0,
\eea
where $\xi^a_{ij}$ and $\xi^a_{ijk}$ are fully symmetric, traceless
tensors with the detailed forms given in Ref. \cite{murakami2004}.

The five rank-2 tensors are often denoted as spin-nematic matrices.
They are
\bea
&&
\Gamma^1= \frac{1}{\sqrt 3} ( F_x F_y +F_y F_x ), \ \ \
\Gamma^2= \frac{1}{\sqrt 3} ( F_z F_x +F_x F_z ), \ \ \
\Gamma^3= \frac{1}{\sqrt 3} ( F_z F_y +F_y F_z ), \ \ \ 
\nonumber \\
&& 
\Gamma^4= \frac{1}{\sqrt 3} ( F_z^2-\frac{5}{4} ), 
\hspace{13mm}
\Gamma^5= \frac{1}{\sqrt 3} ( F_x^2 -F_y^2 ).
\eea
Remarkably, for $F=\frac{3}{2}$, they anticommute with each other 
and form a basis of the Dirac $\Gamma$ matrices as
$\{\Gamma_a, \Gamma_b\}=2\delta_{ab}$, i.e., they form an $SO(5)$
vector.
More explicitly, they are expressed as
\bea
\Gamma^1=\left (
\begin{array} {cc}
0 & -i I\\
i I& 0
\end{array} \right) , \ \ \ 
\Gamma^{2,3,4}=\left ( \begin{array}{cc}
{\vec \sigma}& 0\\
0& {-\vec \sigma} \end{array}\right), \ \ \
\Gamma^5=\left( \begin{array} {cc}
0& I \\
I & 0 \end{array} \right ), 
\eea
where $I$ and
$\vec{\sigma}$ are the 2$\times$ 2 unit and  Pauli matrices.
On the other hand, three spin operators $F_{x,y,z}$ and
seven rank-3 spin tensors $\xi^L_{ijk} F_i F_j F_k, (L=1,..,7)$
together form the ten $SO(5)$ generators.
By linear combinations, these generators can be organized into
\bea
\Gamma^{ab}=
-\frac{i}{2} [ \Gamma^a, \Gamma^b] \ \ \ (1\le a,b\le5).
\eea
Consequently, the 16 bilinear operators can also be classified according to
their properties under the $SO(5)$ transformations as
scalar $n$ (density), vector $n_a$ (spin nematics), 
and anti-symmetric tensors $L_{ab}$ (spin and spin cubic tensors):
\bea 
n({\vec  r})&=& \psi^\dagger_\alpha({\vec r})
\psi_\alpha({\vec r}), \ \ \
n_a({\vec r} )= \frac{1}{2}
\psi^\dagger_\alpha({\vec r}) \Gamma^a_{\alpha\beta}
\psi_\beta({\vec r}), \nonumber \\
L_{ab}({\vec  r})&=&  -\frac{1}{2}\psi^\dagger_\alpha({\vec r})
\Gamma^{ab}_{\alpha\beta} \psi_\beta({\vec r}). 
\label{ch3:phbilinear}
\eea
$L_{ab}$ and $n_a$ together form the $SU(4)$, or isomorphically,
the $SO(6)$ generators.
The $n$, $n_a$, and $L_{ab}$ in the lattice model
can also be defined accordingly.

Next we study the particle-particle channel bilinears.
Pairing operators can be organized as
$SO(5)$ scalar and vector operators through the
charge conjugation matrix $R=\Gamma_1\Gamma_3$ as
\bea
\eta^\dagger(\vec r)&=&\Re \eta+ i~\Im \eta=
\frac{1}{2} \psi^\dagger_\alpha(\vec r)  R_{\alpha\beta}
\psi^\dagger_\beta(\vec r),\nonumber\\
\chi^\dagger_a({\vec r})&=& \Re \chi_a + i~\Im \chi_a=
 -\frac{i}{2}
\psi^\dagger_\alpha({\vec r}) (\Gamma^a R)_{\alpha\beta}
\psi^\dagger_\beta ({\vec r}). 
\label{eq:pairopt}
\eea
The quintet pairing operators $\chi^\dagger_a$ are just the 
polar combinations of $F_z$'s eigenoperators 
$P^\dagger_{2,m}$ with the relation 
\bea
\hspace{-10mm}
P^\dagger_{0,0}&=&  -\frac{\eta^\dagger}{\sqrt 2}, \ \ 
P^\dagger_{2,\pm2}= \frac{\mp\chi^\dagger_1+ i\chi^\dagger_5}{2},
\ \ \
P^\dagger_{2,\pm1}= \frac{-\chi^\dagger_3\pm i\chi^\dagger_2}{2},\ \ \
P^\dagger_{2,0}= -i \frac{\chi^\dagger_4}{\sqrt 2}.
\eea
The existence of the $R$ matrix is related to the pseudoreality 
of $SO(5)$'s spinor representation.
It satisfies 
\bea
R^2=-1,\ \ \ R^\dagger=R^{-1}=~^t R=-R, \ \ \ R \Gamma^a
R=-^t\Gamma^a, \ \ \ R \Gamma^{ab} R=~ ^t\Gamma^{ab},
\eea 
where $^t\Gamma^{a, ab}$ are the transposed matrices of $\Gamma^{a,ab}$.
The anti-unitary time-reversal
transformation can be expressed as
$T=R~ C$,
where $C$ denotes
complex conjugation and $T^2=-1$. $N$, $n_a$, and $L_{ab}$
transform differently under the $T$ transformation 
\bea 
T n T^{-1} =n,~~  T n_a T^{-1} =n_a,~~ T L_{ab} T^{-1}=-L_{ab}. 
\label{eq:evenodd}
\eea
Under particle-hole transformation, fermions transform as
$\psi_\alpha \rightarrow R_{\alpha\beta} \psi^\dagger_\beta, 
 \psi^\dagger_\alpha \rightarrow R_{\alpha\beta} \psi_\beta.
$
Correspondingly, $L_{ab}, n_a$ transform as
\bea
\psi^\dagger_\alpha L_{ab,\alpha\beta}\psi_\beta \rightarrow
\psi^\dagger_\alpha L_{ab,\alpha\beta}\psi_\beta, \ \ \
\psi^\dagger_\alpha n_{a,\alpha\beta}\psi_\beta \rightarrow
-\psi^\dagger_\alpha n_{a,\alpha\beta}\psi_\beta.
\eea

With the above preparation, the hidden $SO(5)$ symmetry becomes manifest.
We can explicitly check all the ten $SO(5)$ generators 
$L_{ab}$ operators commute with the Hamiltonian
Eq. \ref{eq:ctnHam} and Eq. \ref{eq:latHam}.
In other words, the seven rank-3 spin tensor operators
are hidden conserved quantities,
which play the same role to the Runge-Lentz vectors in the Hydrogen atom.
The kinetic energy parts in Eq. \ref{eq:ctnHam} and Eq. \ref{eq:latHam}
have an explicit $SU(4)$ symmetry.
The singlet and quintet interactions are proportional to
$\eta^\dagger({\vec r}) \eta({\vec r})$ and $\chi_a^\dagger({\vec r})
\chi_a({\vec r})$ respectively, thus reducing the symmetry group from
$SU(4)$ to $SO(5)$. 
When $g_0=g_2$ or $U_0=U_2$, the $SU(4)$
symmetry is restored because $\chi^\dagger_a,\eta^\dagger$
together form its 6 dimensional antisymmetric tensor
representation. In the continuum model, interactions in other even
partial wave channels also keep the $SO(5)$ symmetry. The odd
partial wave scattering include spin 1 and 3 channel interactions
$g_1$ and $g_3$, which together could form the 10-d adjoint representation
of $SO(5)$ at  $g_1 = g_3$.
However, to the leading order,
$p$-wave scattering is weak for neutral atoms, and can thus be safely 
neglected.
In the lattice model, this corresponds to the off-site interactions,
and can be also be neglected for neutral atoms.
The proof of $SO(5)$ invariance in the continuum model applies equally well
in the lattice model at any lattice topology and at any filling level.
For later convenience, we rewrite the lattice Hamiltonian in the
following form:
\bea 
H_0&=& -t \sum_{\langle i,j \rangle}\Big \{ \psi^\dagger(i) \psi(j)
+ h.c.\Big\} \nonumber \\
H_I&=&- \sum_{i, 1\le a\le 5}\Big \{
\frac{V}{2} (n(i)-2)^2+\frac{W}{2} n_a^2(i) \Big \}
- (\mu -\mu_0) \sum_i n(i),
\label{eq:latHam2}
\eea 
with $V=(3 U_0+5 U_2)/ 8$ and $W=(U_2-U_0)/2$.

It is natural to think about the symmetry properties in 
other high fermionic systems with $F=n-\frac{1}{2}~(n\ge 3)$.
Indeed, we can generalize this $Sp(4)$ symmetry to
the $Sp(2n)$ symmetry under certain 
conditions, but fine tuning is generally speaking needed.
In such systems, the $s$-wave channel interactions can be classified
into channels with total spin $S_T=0,2, .., 2n-2$ with coupling
constants $g_0, g_2, ..., g_{2n-2}$, respectively.
The $Sp(2n)$ symmetry appears when $g_2=g_4=...=g_{2n-2}$.
The proof of the $Sp(2n)$ symmetry and an introduction to the
$Sp(2n)$ algebra are given in Ref. \cite{wu2005}.

\subsection{The $SO(8)$ structure in the bipartite lattice systems}
We next further  explore the full symmetry of the spin-$3/2$ 
system by extending the above algebra to include the $p$-$p$ channel.
In spin-1/2 bipartite lattice systems, it is well known  
that the particle density and $\eta$-pairing operators form a
pseudospin $SU(2)$ algebra \cite{yang1990}.
For spin-$3/2$ systems,
we embed the above $SO(5)$ algebra inside the $SO(8)$
algebra  involving both $p$-$h$ and $p$-$p$ channels.
$SO(8)$ is the largest Lie algebra which can be constructed with four 
component fermions.
The $SO(8)$ symmetry is shown to be dynamically generated in
spin-$\frac{1}{2}$ two-leg ladder systems, and its effect
was extensively studied \cite{lin1998,konik2001}.
The $SO(8)$ algebra here is attributed with
different physical meaning.
Its generators $M_{ab}~(0\le a<b\le 7)$ reads
\bea\label{so8algebra}
M_{ab}=
\left( \begin{array}{cccc}
0&  \Re \chi_1~ \sim~ \Re \chi_5 & N  &   \Re \eta  \\
 &                                  &\Im \chi_1 &n_1  \\
 &        L_{ab}                    & \sim      &\sim \\
 &                                  &\Im\chi_5  &n_5 \\
 &                                  &   0       &-\Im\eta \\
 &                                  &           &0 \\
\end{array} \right), \nonumber
\eea
with $N =(n-2)/2$.
In the bipartite lattice, we define the global $SO(8)$ generators
as $M_{ab}=\sum_i M_{ab}(i)$ for $M_{ab}=n, n_a, L_{ab}$ in the
$p$-$h$ channel,
and $M_{ab}=\sum_i (-)^i M_{ab}(i)$ for $M_{ab}= \Re \eta,
\Im \eta, \Re \chi_a, \Im \chi_a$ in the $p$-$p$ channel.

At $U_0=5~U_2$ and $\mu=\mu_0~ (\mu_0=\frac{U_0+5 U_2}{4})$,
the three spin singlet sets of $E_{1,4,6}$ become degenerate. 
They form a spin-1 representation for the $SU(2)$ algebra spanned by
$\eta^\dagger,\eta, N$.
In this case,  $H_I$ can be rewritten as 
\bea
H_I=\sum_{i, 1\le a,b \le 5} \{-U_2~L_{ab}^2(i)-(\mu-\mu_0) n(i)\},
\eea
by using the Fierz identity 
$
\sum_{1\le a  \le 5} L^2_{ab}(i)
+\sum_{1\le a \le 5} n_a^2(i)+5 N^2 (i)=5.
$
The symmetry at half-filling is $SO(5)$ $\otimes$ $SU(2)$, which unifies the
charge density wave (CDW) and the singlet pairing (SP) order parameters.
Away from half-filling, this symmetry is broken
but $ \eta,\eta^\dagger$ are still eigen-operators since
$ [H, \eta^\dagger] = -(\mu-\mu_{0}) \eta^\dagger$, and $[H, \eta] =
(\mu-\mu_{0}) \eta$.

Similarly, at $U_0=-3U_2$ and $\mu=\mu_0$, the two spin singlet states
$E_{1,6}$ and the spin quintet states $E_3$ become degenerate.
They form a 7-vector representation for the $SO(7)$ group spanned
by $M_{ab} (0\le a < b\le 6)$.
The $H_I$ can be reorganized into
\bea
H_I=\sum_{i,0\le a <b \le 6} \Big \{ \frac {2}{3} U_2 ~
M_{ab}(i)^2 -(\mu-\mu_0) n(i)\Big \}.
\eea
The $SO(7)$ symmetry is exact at half-filling. 
Its 7-d vector representation unifies the order parameters of
the staggered $\sum_i (-)^i n_a(i)$,  and the singlet pairing
$\sum_i \eta^\dagger(i)$. 
Its 21-d adjoint representation unifies the order parameters of
the staggered $\sum_i (-)^i L_{ab}(i)$, the CDW 
$\sum_i (-)^i N(i)$, and quintet pairing $\sum_i \chi^\dagger (i)$.
Away from half-filling, quintet pairing
operators are spin 2 quasi-Goldstone operators
$[H,\chi^\dagger_{a}(\chi_a) ]=\mp(\mu-\mu_{0}) \chi^\dagger_{a}(\chi_a)$.
These $\chi$-modes are just the analogs of the $\pi$ modes in the high T$_c$
context\cite{zhang1997}.

%******************************************************************
\section{Effects of the $SO(5)$ symmetry}
\label{sect:effect}
In this section, we review the $SO(5)$ effect to the
Fermi liquid theory \cite{wu2003}, the mean field
phase diagram \cite{wu2003}, the sign problem in quantum
Monte-Carlo simulations \cite{wu2003,wu2005}. 
We also construct effective strong coupling models at half-filling 
in Sect. \ref{subsect:strong}.

\subsection{The $SO(5)$ Fermi liquid theory}
The Fermi liquid theory in spin $3/2$ systems is 
simplified by the $SO(5)$ symmetry.
A particle-hole pair in such systems can carry total spins with $S_T=0,1,2,3$,
thus it generally requires four Fermi liquid functions in these channels.
However,  the $SO(5)$ symmetry reduces them to three independent sets as
\bea
f_{\alpha\beta,\gamma\delta}(\vec p, \vec p^\prime)&=& 
f_s(\vec p, \vec p^\prime)+
f_v(\vec p, \vec p^\prime) (\frac{\Gamma^a}{2})_{\alpha\beta}
(\frac{\Gamma^a}{2})_{\gamma\delta} 
+f_t(\vec p, \vec p^\prime) (\frac{\Gamma^{ab}}{2})_{\alpha\beta}
(\frac{\Gamma^{ab}}{2})_{\gamma\delta}, \ \ \  \ \ \
\eea 
where $\vec p$ and $\vec p\prime$ are the momenta of two particles near the
Fermi surface, $f_{s,v,t}$ describe the interaction in
the $SO(5)$ scalar, vector, and tensor channels, respectively.
Within the $s$-wave scattering approximation, these
functions become constants as $f_s= (g_0+5 g_2)/16$,
$f_v= (g_0-3 g_2)/4$, $f_t= -(g_0+ g_2)/4$.
In other words, the two channels of $S_T=1,3$ are degenerate and
included in the tensor channel.
The susceptibility in each channel reads
$\chi_{s,v,t}({\vec r}-{\vec r}^\prime, t-t^\prime)=
-i \theta(t-t^\prime) \avg{|[\hat{O}_{s,v,t}({\vec r},t), 
\hat{O}_{s,v,t}({\vec r^\prime},t^\prime)]|} 
$
with $\hat{O}_{s,v,t}=n$, $n_a$, $L_{ab}$, respectively.
Within the  random-phase approximation, the Fourier transforms in the
momentum and frequency space are
$\chi_{s,v,t}(q,\omega)=\chi^0(q,\omega)/[ 1+ f_{s,v,t} \chi^0(q,\omega)]$,
where $\chi^0(q,\omega)$ is the standard 
Lindhard function of the free systems.
The degeneracy in the spin 1 and 3 channels was first pointed out 
in Ref. \cite{yip1999}, but it appears accidentally there.
It is actually  exact and protected by the generic $SO(5)$ symmetry.
Experiments in the Fermi liquid
regime can determine the four Fermi liquid constants in the
$S_T=0,1,2,3$ channels separately and verify the degeneracy
between spin 1 and 3 channels.

\subsection{Mean field analysis at half-filling in bipartite lattices}
\begin{figure}
\centering\epsfig{file=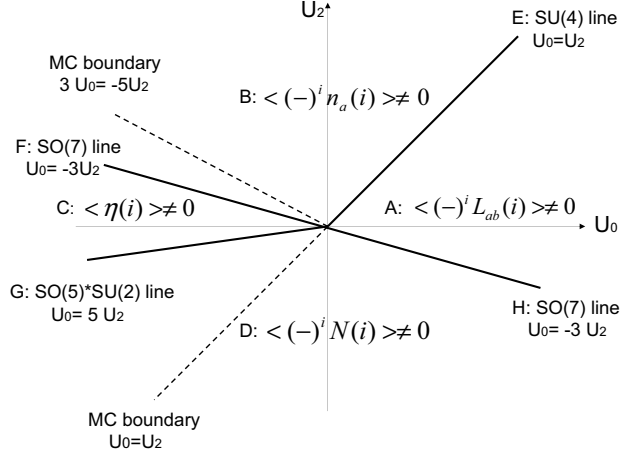,clip=1,width=0.6\linewidth,angle=90}
\caption
{The MF phase diagram at half-filling in a bipartite lattice.
A) and B): staggered phases of the $SO(5)$ adjoint and vector Reps;
C): the singlet superconductivity; D): CDW;
E), F), G) and H): exact phase boundaries with higher symmetries.
Between the dashed lines ($U_0\le U_2\le -3/5 U_0$) ,
a Monte-Carlo algorithm free of the sign problem is possible.
From Wu {\it et al.} Ref. [21].
%\cite{wu2003}.
}\label{fig:mfphase} 
\end{figure}

In the weak coupling limit, we perform a mean-field analysis to the lattice
Hamiltonian Eq. \ref{eq:latHam} at half-filling in a bipartite lattice.
The decoupling is performed in all of the direct, exchange, and
pairing channels as
\bea
H_{MF}&=& -t\sum_{\langle ij,\sigma\rangle}
\Big \{ c^\dagger_{i,\sigma} c_{j,\sigma}+h.c. \Big \}
-(\mu-\mu_0) \sum_{i,\sigma}  n(i)
- \frac{3 U_2-U_0}{2} \sum_{i,1\le a \le 5} \avg{n_a(i} n_a(i) \nonumber\\
&-& \frac{ U_0+U_2}{ 2}
\sum_{i, 1\le a<b \le 5} \avg{L_{ab}(i)} L_{ab}(i)
+\frac{U_0+5U_2}{2} \sum_i \avg {N(i)} N(i) \nonumber \\
&+& U_0\sum_i \Big \{
\avg{ \Re \eta(i) }~\Re \eta(i) +\avg{\Im \eta(i)} ~\Im \eta(i)
\Big \} 
+U_2 \sum_{i,1\le a\le 5} \Big \{
\avg { \Re\chi_a (i)}  \Re\chi_a(i) \nn \\
&+&\avg { \Im\chi_a (i)}  \Im\chi_a(i)
\Big \} 
\label{eq:meanfield}
\eea 

We solve the Eq. \ref{eq:meanfield} self-consistently at
half-filling in the 2D square lattice with the phase diagram 
depicted in Fig. \ref{fig:mfphase}. 
We use the following MF ansatz:
\bea
\avg {n_a(i)}&=&(-)^i \overline{n}_a , \ \ \
\avg{N(i)}=(-)^i \overline{N}, \ \ \
\avg {L_{ab}(i)} =(-)^i \overline {L}_{ab} \nonumber \\
\avg{\eta(i)}&=&\overline \eta, ~~~~~~~~~~ 
\avg {\chi_{a}(i)}=\overline \chi_a.
\eea
There are four phases A, B, C and D with bulk area
separated by high symmetry lines E, F, G and H.
In phase A, the staggered order parameters $\avg{(-)^i L_{ab}(i)\neq 0}$
form $SO(5)$'s adjoint representation (Rep) with the residue symmetry 
$SO(3)\otimes SO(2)$, thus the Goldstone (GS) manifold is
$SO(5)/[SO(3)\otimes SO(2)]$.
In phase B, the staggered order parameters
$\avg{(-)^i n^a(i)}\neq 0$ form $SO(5)$'s vector Rep with
the residue symmetry $SO(4)$ and the GS manifold  $S^4$.
In phase C and D, order parameters are the singlet superconductivity 
$\avg {\eta(i)}\neq 0$ and charge density wave (CDW) 
$\avg{(-)^i N(i)}\neq 0$ respectively,  where the $SO(5)$ 
symmetry is unbroken.
$SU(4)$ symmetry arises at line E where the staggered
order parameters $(-)^i L_{ab}(i)$ and $(-)^i n^a(i)$ become 
degenerate with GS manifold $U(4)/[U(2)\otimes U(2)]$. 
Line F and H are characterized by the $SO(7)$ symmetry.
On the line F, order parameters $(-)^i n^a(i)$ and $\eta(i)$ 
are degenerate with the GS manifold $S^6$.
One the line E, order parameters $(-)^i L_{ab}(i)$, CDW and
$\chi^a(i)$ are degenerate with the GS manifold $SO(7)/[SO(5)
\otimes SO(2)$.
The $SO(5)\otimes SU(2)$ symmetry arises at line G,
where CDW and singlet pairing become degenerate.
Order parameters in each phase and corresponding GS modes are
summarized in Table 3.1.

\begin{table}
\begin{center}
\begin{tabular}{|c|l|c|c|c|}   \hline
Phase &  order parameters  &  GS manifold & GS modes \\ \hline
A& $(-)^i L_{ab}(i)$&$SO(5)/[ SO(3) \otimes SO(2)]$&6 \\ \hline
B& $ (-)^i n_a(i)$  & $SO(5)/ SO(4)\equiv S^4$ & 4 \\  \hline
C& $ \eta(i)$       & $U(1)$&1  \\ \hline
D& CDW             & /  &/  \\ \hline\hline
E& $ (-)^i n_a(i),  (-)^i L_{ab}(i)$&
   $U(4)/[U(2)\otimes U(2)]$&  8 \\ \hline
F& $ (-)^i n_a(i), \eta(i)$& $SO(7)/SO(6)\equiv S^6$ & 6 \\ \hline
G& CDW, $\eta(i)$ &$SO(3)/SO(2)\equiv S^2$ &2 \\ \hline
H& CDW, $\chi_a(i), (-)^i L_{ab}(i)$&
$SO(7)/[SO(5)\times SO(2)]$& 10  \\ \hline
\end{tabular}
\caption{
Order parameters, the corresponding Goldstone manifolds and the number
of Goldstone modes in each phase on the bipartite lattice at half-filling.
From Wu {\it et al.} Ref. [21].
%\cite{wu2003}.
}
\end{center}
\label{tb:gs}
\end{table}

\subsection{Strong coupling modes at half-filling}
\label{subsect:strong}
\begin{figure}
\centering\epsfig{file=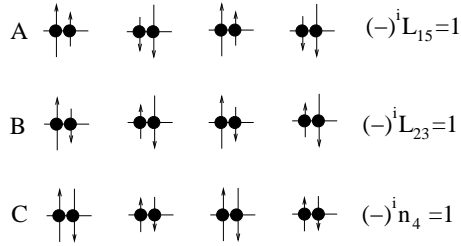,clip=1,width=6cm,angle=0}
\caption{Three configurations of  classical Neel states
on the $SU(4)$ line.
}\label{fig:neel}
\end{figure}

Next we construct the effect Hamiltonians for the bosonic
sector in the strong coupling limit.
Around the $SU(4)$ line E with $U_0=U_2=U$, the low energy states are the
six states with double occupancy $E_{3,4}$.
The exchange model in this projected Hilbert space reads
\bea
H_{eff}&=& 
J_L  \sum_{ 1\le a<b \le 5}  L_{ab}(i) L_{ab}(j)
+J_n \sum_{1\le a \le 5} n_a(i) n_a(j)  
+ \frac{\Delta U_1}{4} \sum_{i,1\le a<b \le 5} L^2_{ab}(i), \ \ \ \ \ \
\label{eq:heisenberg}
\eea
with $\Delta U_1= U_2-U_0$.
Along the $SU(4)$ line, $\Delta U_{1}=0$ and $J_L=J_n= 4 \frac{t^2}{U}$,
Eq. \ref{eq:heisenberg} is just the $SU(4)$ Heisenberg model
with each site in the  6-d representation.
At three dimensions, we expect the ground state to be
long range Neel ordered.
Three possible classical configurations  A, B, C are shown in Fig. 
\ref{fig:neel}, which can be rotated into each other under $SU(4)$ operations.
They are eigenstates of three $SU(4)$ Cartan 
generators:  $L_{23}=\frac{1}{2}
(n_{\frac{3}{2}}-n_{\frac{1}{2}}+n_{\frac{-1}{2}}-n_{\frac{-3}{2}})$,
$L_{15}=\frac{1}{2}
(n_{\frac{3}{2}}+n_{\frac{1}{2}}-n_{\frac{-1}{2}}-n_{\frac{-3}{2}})$,
and $n_4=\frac{1}{2}(n_{\frac{3}{2}}-n_{\frac{1}{2}}-
n_{\frac{-1}{2}}+n_{\frac{-3}{2}})$.
A nonzero onsite $\Delta U_1$ term reduces the symmetry down to
$SO(5)$ and brings the difference between $J_L$ and $J_n$.
The latter contribution is at the order of $(t/U)^2$
compared to the former, thus is less important.
In phase $A$, five quintet states become the lowest energy states,
then the $J_n$ and $\Delta U_1$ terms can be neglected, and thus
Eq. \ref{eq:heisenberg} reduces to the $SO(5)$ Heisenberg model.
The Neel states in phase $A$ can be represented in Fig. \ref{fig:neel}
A and B which breaks TR symmetry.
In phase $B$, the singlet state becomes the lowest energy state,
Eq. \ref{fig:neel} reduces to the $SO(5)$ rotor model whose (spin-nematic)
Neel order keeps the TR symmetry as represented in Fig. \ref{fig:neel} C.
A transition from a site singlet phase to the Neel ordered
phase happens at $z J_1\approx \Delta U_1$ where $z$ is the coordination 
number.

Next we look at the $SO(7)$ line $F$ where the lowest energy state 
is the singlet state $E_4$, and the next bosonic states are the
7-fold degenerate sets of $E_{0,3,6}$.
In phase $B$ or $C$ , $E_{0,6}$ becomes higher or lower than $E_{3}$,
respectively.
An anisotropic $SO(7)$ rotor model can be constructed to describe the
above effect:
\bea
H_{eff,2}&=& J_n \sum_{1\le a \le 5} n_a (i) n_a (j)
- \frac{J_\eta}{2} \big\{\eta^\dagger (i) \eta(j)
+ \eta(i) \eta^\dagger(j)\big\}
\nn \\
&+&   \frac{\Delta U_2}{2}\sum_{0\le a<b \le 6} M^2_{ab}(i)
+\frac{\Delta U_3}{2} N^2(i),
\label{eq:7rotor}
\eea 
where $\Delta U_2= \frac{U_2 - U_0}{6}$ and $\Delta U_3
=\frac{U_0 +3 U_2}{2}$.
Along line F, $\Delta U_3=0$ and $J_n=J_\eta$, thus the antiferromagnetic
spin-nematic order and singlet superconductivity become degenerate.
This $SO(7)$ symmetry is the generalization of the $SO(5)$ theory
of high T$_c$ superconductivity \cite{zhang1997}.
In phase C, $\Delta U_3<0$, the staggered
spin-nematic order $n_a$ is disfavored,
and Eq.  \ref{eq:7rotor} reduces to the $U(1)$ rotor model for 
the singlet superfluidity.
On the other hand, for $\Delta U_3>0$ in phase B, Eq.  \ref{eq:7rotor}
reduces to the $SO(5)$ rotor model for $n_a$.

Along line $G$ and in phases $C$ and $D$, the lowest
energy states are three  $SO(5)$ singlets of $E_{1,4,6}$ 
which form a spin-1 Rep for the pseudospin $SU(2)$ algebra.
The effective model is anisotropic spin-1 Heisenberg-like as
\bea
H_{eff,3}= \frac{-J_\eta}{2}  \big\{\eta^\dagger (i) \eta(j)+ \eta(i) 
\eta^\dagger(j) \big\} + J_n N(i) N(j)+\Delta U_4 N^2(i),
\label{eq:psu2}
\eea
with $\Delta U_4= \frac{-U_0+5 U_2}{2}$.
Along the $SO(5)\otimes SU(2)$ line $G$, $J_\eta=J_n=\frac{2 t^2}{|U_0|}$ 
and $\Delta U_4=0$,  the singlet superfluidity and CDW are degenerate.
In the phase $C$ where $\Delta U_4>0$, 
Eq. \ref{eq:psu2} reduces into the $U(1)$ rotor model for superfluidity,
while in the phase $D$ where $\Delta U_4<0$, Eq. \ref{eq:psu2}
reduces into the Ising model for CDW.

Similarly, around the other $SO(7)$ line H, the sets of $E_{1,4,6}$ 
consist the low energy Hilbert space.
We can construct the anisotropic $SO(7)$ Heisenberg model in the 
7-vector representation as
\bea
H_{eff}&=& 
J_L  \sum_{ 1\le a<b \le 5}  L_{ab}(i) L_{ab}(j)
-\frac{J_\chi}{2} \sum_{1\le a \le 5} \big\{\chi^\dagger_a(i) \chi_a(j)  
+ \chi_a(i) \chi^\dagger_a(j) \big\}\nn \\
&+& J_N N(i) N(j)
+ {\Delta U_5} \sum_{i,1\le a<b \le 5} N^2(i),
\label{eq:so7hsbg}
\eea 
where $J_L=J_\chi=J_N= \frac{4t^2}{U_0+U_2}$ and 
$\Delta U_5=0$ around line $H$.
In phase A where $\Delta U_5> 0$, Eq. \ref{eq:so7hsbg} reduces into
the $SO(5)$ Heisenberg model,
while in phase B where $\Delta U_5<0$, it reduces into the
Ising model for the CDW order.

\subsection{Quantum Monte-Carlo sign problem in spin-$3/2$ 
systems}
The sign problem is a major difficulty for the quantum Monte-Carlo 
(QMC) simulations to apply to fermionic systems.
In spin-$3/2$ systems, due to the structure of $SO(5)$ algebra,
the sign problem in the the auxiliary field QMC \cite{BLANKENBECLER1981}
is shown to be absent in a large part of the phase diagram
\cite{wu2003,capponi2004,wu2005}.

We use the equivalent version  of the lattice 
Hamiltonian Eq. \ref{eq:latHam2} where interactions are represented
in terms of the TR even operators $n$ and $n_a$.
By using the Hubbard-Stratonovich (H-S) transformation, the
four-fermion interactions of $n^2$ and $n_a^2$ are decoupled at $V, W>0$
(or  $-3/5~U_0>U_2>U_0$), and the resulting
partition function can be written as 
\bea
Z=
\exp\Big\{ -V \int_0^\beta d\tau \sum_i (n(i,\tau)-2)^2
-W \int_0^\beta d\tau \sum_{i,a} n_a^2(i,\tau) \Big \} 
\det \Big \{ I+ B \Big \}. \ \ \
\eea
The functional determinant
$I+B= I+ {\cal T} e^{-\int_0^\beta d\tau \nonumber
H_K +H_i(\tau)}$ is from the integration of fermion 
fields, $n$ and $n_a$ are real H-S bose fields.
The kinetic energy $H_K$ and
time-dependent decoupled interaction $H_i(\tau)$ read
\bea
H_K&=& -t\sum_i (\psi^\dagger_{i,\sigma} \psi_{j,\sigma}+h.c.) 
-\sum_i(\mu-\mu_0) \psi^\dagger_{i\sigma} \psi_{i\sigma}\nonumber \\
H_{int}(\tau)&=& -V\sum_i \psi_{i,\sigma}(\tau) \psi_{i,\sigma}(\tau)
n(i,\tau) -W \sum_{i,a} \psi^\dagger_{i,\sigma}(\tau) 
\Gamma^a_{\sigma,\sigma^\prime} \psi_{i,\sigma^\prime}(\tau) n_a(i,\tau).
\ \ \
\eea

Generally speaking, $\det(I+B)$ may not be positive definite, 
making it difficult to use the probability interpretation
of the functional integrand in the QMC simulation.
However, if the above decoupling scheme satisfies an TR-like symmetry,
the following theorem states the positivity of $\det (I+B)$
\cite{koonin1997,hands2000,wu2005}:
if there exists an anti-unitary operator $T$, such that
\bea
T H_K T^{-1} = H_K,\ \ \ T H_I T^{-1} = H_I, \ \ \ T^2=-1,
\label{eq:theorem}
\eea
then the eigenvalues of the $I+B$ matrix always appear in complex
conjugate pairs no matter $I+B$ is diagonalizable or not, 
{\it i.e.}, if $\lambda_i$ is an eigenvalue, then
$\lambda^*_i$ is also an eigenvalue. If $\lambda_i$ is real, it is
two-fold degenerate. 
In this case, the
fermion determinant is positive definite,
\bea
\det (I+B) = \prod_i |\lambda_i|^2 \geq 0.
\eea
The detailed proof can be found in Ref. \cite{wu2005}.
The $T$ operator does not need to be the physical TR operator.
As long as Eq. \ref{eq:theorem} is satisfied,
the positivity of $\det (I+B)$ is guaranteed.
We emphasis that the above criterion applies for any lattice
geometry and doping levels.

Because both the density $n$ and spin-nematic operators $n_a$
are TR even operators, the sign problem disappears as long as
$V, W>0$ or $-3/5~U_0>U_2>U_0$ (see Fig. \ref{fig:mfphase}).
This region includes interesting competing orders such as
the staggered spin-nematic order $n_a$, singlet superfluidity $\eta$,
CDW, and also the phase boundaries of the $SO(7)$ line F
and the $SO(5)\otimes SU(2)$ line G.
The absence of the sign problem provide a good opportunity 
to study the competition among these orders,
in particular, then doping effect, the frustration on the triangular 
lattice, {\it etc}, which are
difficult at low temperatures for previous Monte-Carlo works.

The above sign problem criterion can also be applied to
other systems, such as multi-band Hubbard models.
For example, Capponi {\it et al.} showed a ground state staggered current
order by QMC simulations without the sign problem
in a bi-layer extended Hubbard model \cite{capponi2004}.

%******************************************************************88
\section{Quintet Cooper pairing and half-quantum vortices}
\label{sect:quintet}
When the interaction constant in the quintet channel $g_2$ goes negative,
spin $3/2$ systems can support quintet Cooper pairing, i.e.,
Cooper pairs with total spin 2.
In this section, we discuss the collective modes in this state,
and the topological defect of half-quantum vortex and related
non-Abelian Cheshire charges \cite{wu2005b}.

\subsection{Collective modes in the quintet pairing states}
We write the BCS mean field Hamiltonian for the quintet Cooper pairing
state as
\bea
H_{MF}&=& \int d^D r \Big\{ \sum_{\alpha=\pm
\frac{3}{2}, \pm \frac{1}{2}}
 \psi^\dagger_\alpha({ r})
\big (\frac{-\hbar^2\nabla^2}{2M}-\mu\big) \psi_\alpha({ r}) 
+\sum_{a=1\sim 5} \chi^\dagger_a(r) \Delta_a(r) +h.c.\nn \\
&-&\frac{1}{g_2} \Delta_a^*(r)\Delta_a(r)\Big\}.
\label{eq:qntpair} 
\eea 
$\Delta_a$ is  proportional to the
ground state expectation value of the quintet pairing operators
$\chi_a$ by 
$\Delta_a(r)= g_2\avg{\chi^\dagger_a(r)}~(a=1\sim 5).$
The five $\chi_a$ operators defined in Eq. \ref{eq:pairopt}
are the spin channel counterparts of the five atomic 
$d$-orbitals ($d_{xy}, d_{xz},d_{yz}, d_{3z^2-r^2}, d_{x^2-y^2}$).
They transform as a 5-vector under the $SO(5)$ group. 

The Ginzburg-Landau (GL) free energy around $T_c$ for the quintet pairing
was studied in Ref. \cite{ho1999} without noticing the hidden $SO(5)$
symmetry.
Facilitated by this high symmetry, we organize the GL free
energy in an explicitly $SO(5)$ invariant way,  and 
give a general analysis to the quintet pairing structures.
The GL free energy reads
\bea
F_{GL}&=&\int d^D r~~
\gamma \nabla \Delta^*_a \nabla \Delta_a  + \alpha(T)  \Delta^*_a  \Delta_a
+\frac{\beta_1}{2}  |\Delta^*_a  \Delta_a|^2
+ \frac{\beta_2}{2} \sum_{a<b} 
| \frac{\Delta^*_a \Delta_b- \Delta_b^* \Delta_a}{\sqrt 2}|^2, \nn \\
\alpha&=& -\frac{1}{2} \frac{dn}{d\epsilon} (1- \frac{T}{T_c}),\ \ \
\beta_1=\beta_2= \frac{1}{2} \frac{dn}{d\epsilon} 
\frac{7 \zeta(3)}{8 \pi^2  T_c^2}, \ \ \
\gamma= \frac{n \hbar^2}{4M} \frac{7 \zeta(3)}{8 \pi^2  T_c^2}.
\eea
where $n$ is the particle density, 
$dn/d \epsilon$ is the density of states at the Fermi level, 
and $\zeta(x)$ is the Riemann zeta function.
At $\beta_2>0$ or $\beta_2<0$, two different pairing states are favorable
to minimize the free energy,
i.e., the unitary polar state and non-unitary ferromagnetic state respectively.
The polar state is parameterized as $\Delta_a= |\Delta(T)| e^{i\theta} d_a$, 
where $\theta$ is the $U(1)$ phase, and $\hat d=d_a \hat e_a$ is a 
5-$d$ unit vector in the internal spin space.
The ferromagnetic pairing state is described as
$\Delta_a= |\Delta(T)| e^{i\theta} ( d_{1a} + i~  d_{2a})$
with $\hat d_{1}$, $\hat d_2$ two orthogonal 5-d unit vectors.
The standard Gor'kov expansion gives $\beta_2>0$,
thus the polar state is stable in the framework of the BCS theory.
In the following, we focus on  the polar state
which is also expected to be stable at zero temperature.

At $T=0K$, various low energy dissipationless Goldstone 
modes exist due to the broken symmetries.
In addition to the usual phonon mode,
four branches of spin wave modes carrying $S=2$ arise from the breaking
of the $SO(5)$ symmetry down to $SO(4)$.
The above GL free energy constructed around $T_c$
does not apply to describe these Goldstone modes.
For this purpose, Cooper pairs can be treated as
composite bosons. This treatment gives a good description 
of the phonon mode in the neutral singlet BCS superfluid
in Ref. \cite{aitchison1995,stone1995}.
Here, we generalize it to the quintet pairing
by considering a phenomenological Hamiltonian
for spin 2 bosons
\bea
H_{eff}&=&\int d^D r ~\frac{\hbar^2}{4 M} \sum_{ a}
|\nabla \Psi_a|^2
+\frac{1}{2\chi_\rho} (\Psi^*_a \Psi_a 
-\rho_0)^2
+\frac{1}{2\chi_{sp}} \sum_{a<b}(\Psi^*_c L^{ab}_{cd} \Psi_d)^2
, \ \ \ \ \ \
\label{eq:effham}
\eea
where $\Psi_{a}$s are the boson operators in the polar basis,
$L^{ab}_{cd}=i(\delta_{ac}\delta_{bd}-\delta_{ad}\delta_{bc})$ are
the $SO(5)$ generators in the $5\times 5$ representation,
the equilibrium Cooper pair density $\rho_0$ is half of 
the particle density $\rho_f$,
$\chi_\rho$ and $\chi_{sp}$ are proportional to the compressibility
and $SO(5)$ spin susceptibility respectively.
Taking into account the Fermi liquid correction,
$\chi_\rho=N_f/[4(1+F_0^s)]$ and $\chi_{sp}=N_f/[4(1+F_0^t)]$,
where $N_f$ is the fermion density of states at the Fermi energy,
$F^{s,t}_0$ are the Landau parameters defined
in the $SO(5)$ scalar and tensor channels \cite{wu2003} respectively.
We introduce $\rho(r)$ and $l_{ab}(r)$ as the Cooper pair density
and $SO(5)$ spin density respectively, and parameterize
$\Psi_a=\sqrt\rho_0 e^{i\theta} d_a$.
Using the standard commutation rules between $\rho$ and $\theta$,
$l_{ab}$ and  $\hat d_a$, we arrive at
\bea
&&\chi_\rho \partial_t^2 \theta -\frac{\hbar^2\rho_s}{2M} \nabla^2 \theta
=0, \\
&&\partial_t l_{ab}= \frac{\hbar^2\rho_{sp}}{2 M} ( d_a \nabla^2  d_b 
-  d_b \nabla^2  d_a), \ \ \
\chi_{sp}\partial_t  d_a = - l_{ab}  d_b,
\label {EOM1}
\eea
where $\rho_s$ and $\rho_{sp}$ are the superfluid  density and
spin superfluid density respectively.
At $T=0K$ in a Galilean invariant system, $\rho_s$ is just
$\rho_f/2$, while $\rho_{sp}$ receives Fermi liquid
corrections as $\rho_{sp}/\rho_s=(1+F^v_1/3)/(1+ F^s_1/3)$
\cite{leggett1975} where $F_1^v$ is the Landau parameter
in the $SO(5)$ vector channel \cite{wu2003}.
The sound and spin wave velocities are obtained
as $v_s=\sqrt{\rho_0/(2 \chi_\rho M)}$
and $v_{sp}=\sqrt{\rho_0/(2 \chi_{sp} M)}$, respectively.

\subsection{Half-quantum vortex}
\begin{figure}
\centering\epsfig{file=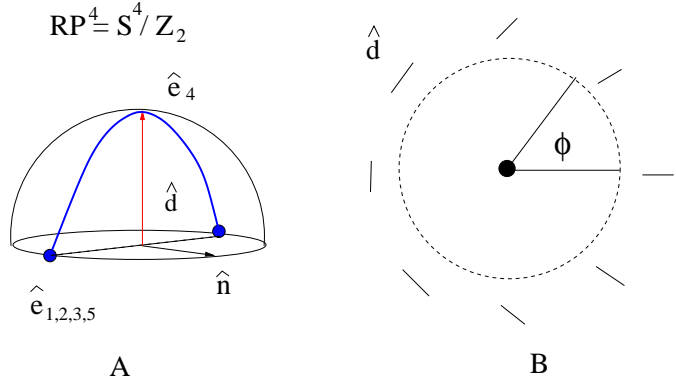,clip=1,width=0.7\linewidth,angle=0}
\caption{A) The Goldstone manifold of $\hat d$ is a 5D
hemisphere $RP^4$. It contains a class of non-contractible loops
as marked by the solid curve. B) The $\pi$-disclination of $\hat
d$ as a HQV. Assume that $\hat d \parallel \hat e_4$  at
$\phi=0$.
As the azimuthal angle $\phi$ goes from 0 to $2\pi$,
$\hat d$ is  rotated at
the angle of $\phi/2$ around any axis $\hat n$ in the $S^3$
equator spanned by $\hat e_{1,2,3,5}$.
From Wu {\it et al.} Ref. [24]. 
%\cite{wu2005b}
} \label{fig:GSManifold}
\end{figure}

Superfluids with internal structure can support half-quantum
vortex (HQV) as a stable topological defect
because the order parameter contains a $Z_2$ gauge symmetry.
For the quintet pairing case, $\Delta_a=|\Delta|
e^{i\theta} \hat d_a$ is invariant under the combined operation
\bea
\hat d \rightarrow -\hat d,  \ \ \  \theta\rightarrow \theta+ \pi.
\eea
As a result, $\hat d$ is actually a directionless director instead of
a true vector.
Thus the fundamental group of the GS manifold is $\pi_1(RP^4\otimes 
U(1))=Z\otimes Z_2$ as depicted in Fig. \ref{fig:GSManifold} A.
The $Z_2$ feature gives rise to the existence of HQV as depicted 
in Fig. \ref{fig:GSManifold} B.
As we move along a loop enclosing HQV, the $\pi$ phase mismatch 
in the $\theta$ field is offset by a $\pi$-disclination in the 
$\hat d$-field, thus $\Delta_a$'s are still  single-valued. 

Next we check the energetics of HQV.
The static energy function can be written as 
\bea 
E=\int d^D r\frac{\hbar^2}{4M} \Big\{ \rho_s (\nabla \theta)^2+ \rho_{sp}
(\nabla \hat d)^2 \Big \}. 
\eea 
The energy density per unit length of a single quantum vortex 
is $E_1=\frac{\hbar^2}{4M} \rho_s \log \frac{L}{a}$, while that of
two isolated HQVs is $E_2=\frac{\hbar^2}{8M} (\rho_s+\rho_{sp}) \log
\frac{L}{a}$.
Thus for $\rho_{sp}<\rho_s$, a single quantum vortex is
energetically less favorable than a pair of HQVs.
Although at the bare level $\rho_s=\rho_{sp}$,
$\rho_{sp}$ receives considerable Fermi liquid correction and
strong reduction due to quantum fluctuations in the 5D internal
space. Generally speaking, $\rho_{sp}<\rho_{s}$ holds
in terms of their renormalized values. 

An interesting property of the HQV is its Alice string behavior
\cite{schwarz1982,bucher1999}.
In this context, it means that a quasi-particle change its spin 
quantum numbers after it adiabatically encircles the HQV.
For example, in the $^3$He-A phase, a quasi-particle with spin 
$\uparrow$ changes to $\downarrow$ up to a $U(1)$ Berry phase.
The HQV in the quintet pairing case behaves as a non-Abelian
generalization of the above effect.
Without loss of generality, we assume that
$\hat d \parallel \hat e_4 $ at the azimuthal angle $\phi=0$.
As $\phi$ changes from $0$ to $2\pi$, $\hat d$ is rotated
at the angle of $\phi/2$ in the plane
spanned by $\hat e_4$ and  $\hat n$, where $\hat n$
is a unit vector perpendicular to $\hat e_4$, i.e., a vector
located in the $S^3$ sphere spanned by $\hat e_{1,2,3,5}$. We
define such a rotation operation as $U(\hat n, \phi/2)$. When $U$
acts on an $SO(5)$ spinor, it takes the form of $U(\hat n,\phi/2)=
\exp\{ -i \frac{\phi}{2} \frac{ n_b \Gamma^{b4}}{2} \}$,
when $U$ acts on an $SO(5)$ vector, it behaves as $U(\hat n,\phi/2)=\exp\{ -i
\frac{\phi}{2} n_b L^{b4} \}$ where $L^{ab}$'s are the $SO(5)$
generators in the $5\times 5$ vector representation.
 The resulting
configuration of $\hat d$ is
$\hat d (\hat n, \phi)=  U(\hat n, \phi/2)
\hat d (\hat n, 0)
= (\cos \frac{\phi}{2} \hat e_{4} -\sin
\frac{\phi}{2} \hat n ).
$
As fermionic quasi-particles circumscribe around the vortex line adiabatically,
at $\phi=2\pi$ fermions with $S_z=\pm\frac{3}{2}$ are rotated into
$S_z=\pm\frac{1}{2}$ and {\it vice versa}.
For convenience, we change the basis $\Psi$ for the fermion wavefunction
to $(|\frac{3}{2}\rangle,|-\frac{3}{2}\rangle,|\frac{1}{2}\rangle,
|-\frac{1}{2}\rangle)^T$.
After taking into account the $\pi$ phase of the superfluid vortex,
$\Psi$ transforms by
\bea
\Psi_a\rightarrow\Psi^{\prime}_a = i U(\hat n,
\pi)_{\alpha\beta} \Psi_{\beta}= \left(
\begin{array}{cc}
0&W\\
W^\dagger&0
\end{array}
\right)_{\alpha\beta} \Psi_\beta
\label{eq:spflip}
\eea
where $W$ is an $SU(2)$ phase depending on the direction of $\hat n$
on the $S^3$ sphere as
\bea
W(\hat n)=\left (\begin{array}{cc}
n_3+in_2&-n_1-in_5\\
n_1-in_5& n_3-in_2
\end{array}
\right). \eea
The non-conservation of spin in this adiabatic
process is not surprising because the $SO(5)$ symmetry is
completely broken in the configuration depicted in
Fig. \ref{fig:GSManifold} B.

\subsection{$SO(4)$ Cheshire charge}
\begin{figure}
\centering\epsfig{file=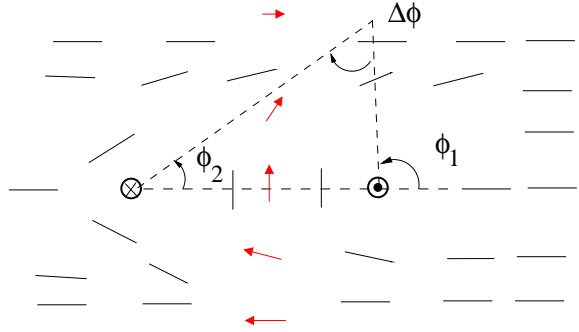,clip=1,width=0.6\linewidth,angle=0}
\caption{The configuration of a $\pi$-disclination pair or loop
described by Eq. \ref{eq:hqvpair}.
$\phi_{1,2}$ and $\Delta \phi$ are azimuthal angles and
$\hat d (\vec r) \parallel \hat e_4$
as $\vec r\rightarrow \infty$.
After a fermion passes the HQV loop, the components with $S_z=\pm\frac{3}{2}$
change to $S_z=\pm\frac{1}{2}$ and {\it vice versa} with
an $SU(2)$ matrix defined in Eq. \ref{eq:spflip}.
From Wu {\it et al.} Ref. [24].
%\cite{wu2005b}
}\label{fig:halfpair}
\end{figure}

Another interesting concept related to the Alice string in the gauge theory
is the Cheshire charge, which means a HQV pair or loop can carry 
spin quantum number.
When a quasi-particle carrying spin penetrate the HQV loop or pair,
the spin conservation is maintain by exciting the Cheshire charge
of the HQV loop or pair.
Below we will construct the $SO(4)$ Cheshire charge in the quintet
Cooper pairing state.

We begin with a uniform ground state where $\hat d$ is parallel to
$\hat e_4$ axis where the $SO(4)$ symmetry generated by
$\Gamma_{ab} (a,b=1,2,3,5)$ is preserved.
This $SO(4)$ algebra can be reorganized into two inter-commutable
$SU(2)$ algebras as
\bea
T_1(T^\prime_1)&=&\frac{1}{4}(\pm\Gamma_{35}-\Gamma_{12}), \ \ \
T_2(T^\prime_2)=\frac{1}{4}(\pm\Gamma_{31}-\Gamma_{25}), 
 \nonumber \\
T_3 (T^\prime_3)&=&\frac{1}{4}(\pm\Gamma_{23}-\Gamma_{15}).
\eea
$T_{1,2,3}$ and $T^\prime_{1,2,3}$ act in the subspaces spanned by $|\pm
\frac{3}{2}\rangle$ and $|\pm \frac{1}{2}\rangle$, respectively.
$SO(4)$ representations are denoted by $|T, T_3; T^{\prime },
T_3^\prime \rangle$, i.e.,  the direct-product of representations
of two $SU(2)$ groups.

Now we construct a HQV loop or pair as depicted in Fig. \ref{fig:halfpair}.
where $\phi_{1,2}$ are azimuthal angles respect to the vortex 
and anti-vortex cores respectively. 
The solution of the configuration of the $\hat d$ vector is described by
the difference between two azimuthal angles
$\Delta\phi=\phi_2-\phi_1$ as
\bea 
\hat d (\hat n, \Delta \phi) =
\cos \frac{\Delta \phi}{2} \hat e_{4} -\sin \frac{\Delta \phi}{2}
\hat n,  
\label{eq:hqvpair} 
\eea 
where $\hat n$ again is a unit vector on the $S^3$ equator. 
This configuration is
called a phase-sharp state denoted as $|\hat n\rangle_{vt}$.
Because the above $SO(4)$ symmetry is only broken within a small
region around the HQV loop, quantum fluctuations of $\hat n$
dynamically restore the $SO(4)$ symmetry as described by the
Hamiltonian 
\bea 
H_{rot}&=& \sum_{a,b=1,2,3,5} \frac{M^2_{ab}}{2
I}, \ \ \  M_{ab}= i (\hat n_a
\partial_{\hat n_b}-\hat n_b \partial_{ \hat n_a}), 
\eea
with the moment of inertial $I= \chi_{sp} \int d^D r ~ \rho_0
\sin^2 \frac{\Delta \phi}{2}$. Thus the zero modes $|\hat
n\rangle_{vort}$ are quantized into the global $SO(4)$ Cheshire
charge state, which is a non-Abelian generalization of the $U(1)$
Cheshire charge in the $^3$He-A phase  \cite{mcgraw1994}. The
Cheshire charge density is localized around the vortex loop. In
contrast, the Cheshire charge in the gauge theory is non-localized
\cite{schwarz1982,bucher1999}.
The HQV loop in the $SO(4)$ Cheshire charge
eigenstates are defined as $ |T T_3; T^\prime
T_3^\prime\rangle_{vt} =\int_{\hat n \in S^3} d \hat n~ F_{T T_3;
T^\prime T_3^\prime} (\hat n) ~|\hat n\rangle_{vt}, $ where $ F_{T
T_3; T^\prime T_3^\prime} (\hat n)$ are the $S^3$ sphere harmonic
functions. 

When a quasiparticle penetrates a HQV loop or pair in the quintet
Cooper pairing state, quantum entanglement is generate between
them in the final state.
We demonstrate this through a concrete example, 
with the initial state $|i\rangle$ made from a zero charged HQV loop 
and a quasiparticle with  $S_z=\frac{3}{2}$ as
\bea |i\rangle &=&\int_{\hat n\in S^3} d\hat n  ~|\hat
n\rangle_{vt} \otimes (u~ c^\dagger_{\frac{3}{2}} +v
~c_{-\frac{3}{2}})|\Omega\rangle_{qp}, \eea where $|\Omega\rangle_{qp}$ is
the vacuum for Bogoliubov particles. 
For each phase-sharp state
$|\hat n\rangle_{vt}$, the particle changes spin according to Eq.
\ref{eq:spflip} in the final state $|f\rangle$. The superposition of
non-Abelian phase gives 
\bea 
|f\rangle &=& \int_{\hat n\in S^3}d
\hat n ~ \Big\{ u ~(W^\dagger_{11} c^\dagger_{\frac{1}{2}}+
W^\dagger_{21}
c^\dagger_{-\frac{1}{2}})  + v~ (W^T_{12} c_{\frac{1}{2}} 
+ W^T_{22} c_{-\frac{1}{2}} )\Big\} 
|\hat n \rangle_{vt} \otimes |\Omega\rangle_{qp}
\nonumber \\
&=&\int_{\hat n\in S^3}d \hat n  (\hat n_3-i \hat
n_2) |\hat n \rangle_{vt} \otimes (u~ c^\dagger_{\frac{1}{2}} +v
~c_{-\frac{1}{2}})|\Omega\rangle_{qp}  \nonumber \\
&-& \int_{\hat n\in S^3}d \hat n (\hat n_1-i \hat
n_5) |\hat n\rangle_{vt} \otimes (u~ c^\dagger_{-\frac{1}{2}} -v
~c_{\frac{1}{2}})|\Omega\rangle_{qp}.
\label{eq:cheshire} 
\eea 
In terms of the $SO(4)$
quantum numbers, $|i\rangle$ is in a product state as
$|00;00\rangle_{vt} \otimes
|\frac{1}{2}\frac{1}{2};00\rangle_{qp}$, and $|f\rangle$ is
\bea
|\frac{1}{2}\frac{1}{2}; \frac{1}{2}\frac{-1}{2}
\rangle_{vt}\otimes |00;\frac{1}{2}\frac{1}{2} \rangle_{qp}-
|\frac{1}{2}\frac{1}{2}; \frac{1}{2}\frac{1}{2}\rangle_{vt}
\otimes |00;\frac{1}{2}\frac{-1}{2}\rangle_{qp}. \ \ \
\label{eq:entag}
\eea
In the channel of the $(T^{\prime },T_3^\prime)$,
the final state is exactly an entangled Einstein-Podolsky-Rosen 
(EPR) pair made up from the HQV loop and the quasi-particle.

%***************************************************************
\section{Quartetting instability in 1D spin-$3/2$ systems}
\label{sect:quartet}
Quartetting instability is a four-fermion counterpart of the
Cooper pairing in spin-$3/2$ systems,
whose order parameter can be written as $O_{qrt}(r)=\
\psi^\dagger_{\frac{3}{2}}(r)
\psi^\dagger_{\frac{1}{2}}(r) \psi^\dagger_{\frac{-1}{2}}(r)
\psi^\dagger_{\frac{-3}{2}}(r)$.
The quartet is an  $SU(4)$ singlet which can be considered
as a four-body EPR state.
This kind of order is very difficult to analyze in high dimensions
because $O_{qrt}$ contains four fermion operators.
It lacks of a Bardeen-Cooper-Schrieffer (BCS) type well-controlled
mean field theory.
Nevertheless, considerable progress has been made in 1D systems.
By  using Bethe ansatz, Scholttmann \cite{schlottmann1994} 
showed that the ground state of the $SU(2N)$ model is characterized
by the formation of the 2N-particle baryon-like bound state.
In contrast, the Cooper pairing can not exist because two particles can
not form an $SU(2N)$ singlet. 
The strong quantum fluctuations in the spin channel in 1D suppress 
the Cooper pairing. 

In this section we will review the quartetting instability in
1D spin-$3/2$ systems including the $SU(4)$
symmetric model as a special case \cite{wu2005a,lecheminant2005}.
We will show that both quartetting and singlet pairing are allowed 
in different parameter regimes, and study their competitions.

\subsection{Renormalization group and Bosonization analysis}
\label{subsect:1dphase}
\begin{figure}
\centering\epsfig{file=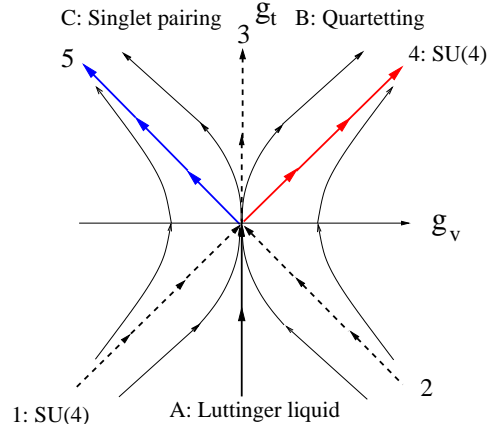,clip=1, width=0.5\linewidth,angle=0} 
\caption
{ RG flows in the parameter space ($g_v, g_t$)
in the spin channel.
Combined with $K_c$ in the charge channel, they
determine various phases as:
A) the gapless Luttinger liquid phase,
B) quartetting phase with QLRO superfluidity at $K_c>2$
or $2k_f$ CDW at $K_c<2$.
C) singlet pairing phase with QLRO superfluidity at $K_c<1/2$
or $4k_f$ CDW at $K_c>1/2$.
They are controlled by the fixed points of $(0,0)$,
$( +\infty, +\infty)$ (line 4), and $(-\infty, +\infty)$ (line 5) 
respectively.
Phase boundaries (line 1, 2, 3) are marked with dashed lines.
From Wu Ref. [28]. 
%\cite{wu2005a}
}\label{fig:RGflow}
\end{figure}

\begin{figure}
\centering\epsfig{file=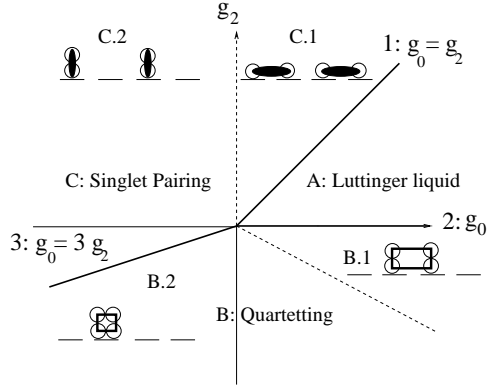,clip=1,width=0.5\linewidth,angle=0}
\caption{ Phase diagram in terms of the singlet and quintet channel 
interaction parameters $g_0$ and $g_2$.
Configurations at quarter filling after the charge gap opens
are shown.
Phase A is the $SU(4)$ gapless spin liquid.
Each of phase B and C splits to two parts.
With $g_u\rightarrow+\infty$, B.1) quartets with both bond and
charge orders, C.1) dimerization of spin Peierls order.
With $g_u\rightarrow-\infty$, CDW phases of B.2) quartets and C.2)
singlet Cooper pairs.
Boundaries among phases A, B and C are marked with solid lines,
and those between phases B.1 and B.2, and C.1 and C.2
are sketched with dashed lines. From Wu Ref. [28].
%\cite{wu2005a}.
}\label{fig:1dphase}
\end{figure}

In 1D systems, we  linearize the spectra around 
the Fermi wavevector $k_f$, and then decompose the
fermion operators into left and
right moving parts as $\psi_{\alpha}=\psi_{R,\alpha} e^{i k_f x}+
\psi_{L,\alpha} e^{-i k_f x}$.
The right moving currents are also classified into $SO(5)$'s scalar, 
vector and tensor currents as
$J_{R}(z) =\psi^\dagger_{R,\alpha}(z) \psi_{R,\alpha}(z), ~
J^a_{R}(z)=\frac{1}{2}\psi^\dagger_{R,\alpha}(z)\Gamma^a_{\alpha\beta}
\psi_{R,\beta}(z)~ (1\le a\le 5), ~
 J^{ab}_{R}(z)=\frac{1}{2}\psi^
\dagger_{R,\alpha}(z)\Gamma^{ab}_{\alpha\beta} \psi_{R,\beta}(z)
~(1\le a<b\le 5),
$
and the left moving currents can be defined accordingly.
The low energy effective Hamiltonian density ${\cal H}={\cal H}_0+{\cal H}
_{int} + {\cal H}^\prime_{int}$ is  written as
\bea\label{ham1}
{\cal H}_0&=& v_f \Big\{ \frac{\pi}{4} J_R J_R +\frac{\pi}{5}
(J^a_R J^a_R +J^{ab}_R J^{ab}_R) +(R\rightarrow L)  \Big\}, \nonumber \\
{\cal H}_{int}&=&   \frac{g_c}{4} J_R J_L + g_v J^a_R J^a_L
 +g_t J^{ab}_R J^{ab}_L,  \nonumber \\
{\cal H}^\prime_{int}&=&   \frac{g^\prime_c}{8} J_R J_R +
 \frac{g^\prime_v}{2} J^a_R J^a_R  +\frac{g^\prime_t}{2} J^{ab}_R J^{ab}_R  
+(R\rightarrow L).
\label{eq:1dHam}
\eea
The Umklapp term is absent in this continuum model,
whose effects become important in the lattice model
at commensurate fillings, and will be discussed in next section.
At the tree level, these dimensionless coupling constants are
expressed in terms of $g_0$, $g_2$ defined above as
\bea
g_c = g_c^\prime = \frac{g_0+ 5 g_2}{2},  \ \ \
g_v= g_v^\prime = \frac{g_0-3g_2}{2},  \ \ \
g_t=g_t^\prime=-\frac{g_0+g_2}{2},
\label{eq:ggpara}
\eea
which are renormalized significantly under the RG process.
The chiral terms in  $H_{int}^\prime$ only renormalize the Fermi velocities,
which  can be neglected at the one loop level. 
At $g_v=g_t$ and $g_v^\prime=g_t^\prime$, or in other words $g_0=g_2$,
the $SU(4)$ symmetry is restored.

The above Hamiltonian Eq. \ref{eq:1dHam} can be bosonized through
the identity 
\bea
\psi_{\alpha R,L} (x)= \eta_\alpha/  \sqrt{2\pi a} \exp \{\pm i \sqrt{\pi} 
(\phi_\alpha(x)\pm \theta_\alpha(x))\} \ \ \ 
(\alpha=\pm\frac{3}{2},\pm\frac{1}{2})
\eea
where the Klein factors
$\eta_\alpha$ are Majorana fermions to ensure the anti-commutation 
relation among fermions of different species.
Boson fields $\phi_\alpha$ and their dual fields $\theta_\alpha$ 
are conveniently reorganized into $\phi_c~(\theta_{c})$ in 
the charge channel, and $\phi_v(\theta_v), \phi_{t1}(\theta_{t1}),
\phi_{t2}(\theta_{t2})$ in the spin channels via
$\phi_{c,v}=\frac{1}{2}(\phi_{\frac{3}{2}}
\pm\phi_{\frac{1}{2}}\pm\phi_{-\frac{1}{2}} +\phi_{-\frac{3}{2}})$, 
$\phi_{t1,t2}=\frac{1}{2}(\phi_{\frac{3}{2}}\mp\phi_{\frac{1}{2}}
\pm\phi_{-\frac{1}{2}}-\phi_{-\frac{3}{2}})$.
Similar expressions hold for $\theta$'s.
In terms of these boson fields, the quadratic part of the Hamiltonian density
is standard  $(\nu=c,v,t_1,t_2)$
\bea 
{\cal H}_0&=&\frac{v_\nu}{2}\sum_\nu \Big\{ K_\nu (\partial_x \theta_\nu)^2 + 
\frac{1}{K_\nu} (\partial_x \phi_\nu)^2 \Big \} \ \ \ \mbox{with} \nn \\
K_{\nu}&=&\sqrt{\frac{2\pi v_f +g^\prime_\nu -g_\nu}
{2\pi v_f+g^\prime_\nu +g_\nu}}, \ \ \
v_\nu=\sqrt{(v_f+\frac{g_\nu^\prime}{2\pi})^2-(\frac{g_\nu}{2\pi})^2},
\eea
where $g_{t1,t2}=g_t, g^\prime_{t1,t2}=g_t^\prime$.
We need to bear in mind that the expression for $K_{\nu}, v_{\nu}$
can not be taken seriously at intermediate and strong couplings
where finite but significant renormalizations to $K_\nu$ and $v_\nu$
take place.
The non-quadratic terms are summarized as 
\bea\label{CosTerms}
{\cal H}_{int}&=&\frac{-1}{2(\pi a)^2} \Big \{\cos\sqrt {4\pi} \phi_{t1}+
\cos\sqrt {4\pi} \phi_{t2} \Big\} 
\Big\{ (g_t+g_v) \cos\sqrt{4\pi}\phi_v \nonumber\\
&+&(g_t-g_v) \cos\sqrt{4\pi}\theta_v
\Big\} - \frac {g_t} { 2(\pi a)^2} 
\cos\sqrt {4\pi} \phi_{t1} \cos\sqrt {4\pi} \phi_{t2},
\eea
with the convention of  Klein factors as $\eta_{\frac{3}{2}} 
\eta_{\frac{1}{2}} \eta_{\frac{-1}{2}} \eta_{\frac{-3}{2}}=1$.
Eq. \ref{CosTerms} contains  cosine terms of both $\phi_v$ and
its dual field $\theta_v$.
Their competition leads to two different spin gap phases
with either $\cos \sqrt{4\pi}\phi_v$ or $\cos \sqrt{4\pi}\theta_v$
pinned as shown below.
The charge channel $c$ remains quadratic.

The renormalization group (RG) equation of $g_{c,v,c}$ can 
be calculated from the standard operator product expansion
technique \cite{senechal1999} as
\bea\label{RG}
&&\frac{ d g_v} {d  \ln (L/a) }= \frac{4}{2\pi}  g_v g_t, \ \ \
\frac{ d g_t}{d \ln (L/a) }= \frac{1}{2\pi} ( 3 g_t^2 + g_v^2 ), \ \ \
\frac{d g_c} {d \ln (L/a)} = 0,
\eea
where $L$ is the length scale and $a$ is the short distance cutoff.
Due to the absence of the Umklapp term, the charge part $g_c$ remains 
unrenormalized  at the one-loop level.
The $SU(4)$ symmetry is preserved in the RG process
along the line $g_v=g_t$.
The RG equations  can be integrated as 
$
|g_t^2-g_v^2|= c |g_v|^{3/2} \ \ \ \mbox{ ($c$: constant)}
$
with the RG flows as shown in Fig \ref{fig:RGflow}.
According to the relation Eq. \ref{eq:ggpara}, the
corresponding boundaries are also shown in  Fig. \ref{fig:1dphase}.

The phase diagram Fig \ref{fig:RGflow} (Fig. \ref{fig:1dphase})
contains three phases at incommensurate fillings.
Phase A is the gapless Luttinger liquid phase lying
in the repulsive interaction regime with $g_2<g_0$.
All the  cosine terms are marginally irrelevant.
The leading order divergent susceptibility are
the $2k_f$-CDW order $O_{2k_f,cdw}=\psi^\dagger_{R\alpha} 
\psi_{L\alpha}$, and
the $2k_f$-SDW orders in the $SO(5)$ vector channel 
$N^a= \psi^\dagger_{R\alpha}(\frac{\Gamma^a}{2})_{\alpha\beta} 
\psi_{L\beta}$, and those in the $SO(5)$ tensor channel 
$N^{ab}= \psi^\dagger_{R\alpha}
(\frac{\Gamma^{ab}}{2})_{\alpha\beta} \psi_{L\beta}$.
All of them are with the scaling dimension of $(K_c+3)/4$.

Phase B is characterized by the formation of quartets.
This phase is controlled by the marginally relevant fixed point
$g_v=g_s\rightarrow +\infty$ (line 4 in Fig. \ref{fig:RGflow}),
and lies in the regime where  attractive interactions dominates
(Fig. \ref{fig:1dphase}).
The spin gap opens with pinned boson fields of
$\phi_v$, $\phi_{t1}$ and $\phi_{t2}$.
The pinned values can be chosen as
$\avg{\phi_v}=\avg{\phi_{t1}}=\avg{\phi_{t2}}=0$ up to a
gauge degree of freedom.
In phase B, every four fermions first form quartets, then quartets
undergo either superfluidity or CDW instabilities.
Because the average distance between two nearest quartets is $d=\pi/k_f$,
the CDW of quartet is of the $2k_f$-type $N$.
Their order parameters reduce to $O_{2k_f,cdw}\propto e^{i\sqrt \pi \phi_c}$
and $O_{qrt}\propto e^{2i\sqrt{\pi} \theta_c}$ respectively.
Checking their scaling dimensions $\Delta_{N}=\frac{1}{4} K_c$
and $\Delta_{qrt}= 1/K_c$,
$O_{qrt}$ wins  over $N$ at  $K_c>2$.
Previous Bethe-ansatz results for the $SU(4)$ case
show the quartets formation in Fig. \ref{fig:RGflow} \cite{schlottmann1994}.
Here, we extend the quartetting regime to the whole phase B.
On the other hand, all the pairing operators $\eta^\dagger$ 
and $\chi^{a\dagger}$ decay exponentially in phase B.

Phase C is characterized by the formation of the singlet Cooper pairs.
The parameter region for this phase is symmetric to phase B
in Fig. \ref{fig:RGflow} under the reflection $g_v\rightarrow -g_v$.
The corresponding fixed point is $-g_v=g_t\rightarrow +\infty$. 
Different from phase B, the dual field $\theta_v$ 
instead of $\phi_v$ is pinned.
Similarly, we can pin the values of bosonic fields
$\avg{\phi_{t1}}=\avg{\phi_{t2}}=\avg{\theta_v}=0$
to minimize the ground state energy.
The two leading competing orders in this phase
are the singlet pairing $\eta^\dagger$ 
and the CDW of pairs.
This CDW is just the  $4k_f$ CDW defined as
$O_{4k_f,cdw}=\psi^\dagger_{R\alpha} \psi^\dagger_{R\beta} 
\psi_{L\beta} \psi_{L\alpha}$.
The expressions of these two orders  reduce into 
$\eta^\dagger\propto e^{-i\sqrt \pi \theta_c}$, and
$O_{4k_f,cdw}\propto e^{\sqrt{4\pi} \phi_c}$.
At $K_c>1/2$, the pairing instability $\eta^\dagger$
wins over the 4$k_f$-CDW.

\subsection{Ising transition between quartetting and pairing phases}
Next we study the competition between the quartetting (phase B)
and singlet Cooper pairing (phase C),
which can be mapped into the phase locking problem of 
two-component superfluidity \cite{leggett1966}.
One component is defined as $\Delta_1^\dagger=\psi^\dagger_{\frac{3}{2}}
\psi^\dagger_{\frac{-3}{2}}$,
and the other one as $\Delta_2^\dagger=\psi^\dagger_{\frac{1}{2}}
\psi^\dagger_{\frac{-1}{2}}$.
Then the singlet pairing operator $\eta^\dagger$ and the quartetting
operators are represented as
\bea
\eta^\dagger=\frac{1}{2}(\Delta^\dagger_1
-\Delta^\dagger_2)\propto  e^{i\sqrt \pi \theta_c }
\cos \sqrt \pi \theta_r, \ \ \
O^\dagger_{qrt}=\Delta^\dagger_1 \Delta^\dagger_1=
e^{i\sqrt{4\pi} \theta_c} \cos 2 \sqrt \pi \phi_v,
\eea
where the charge channel boson field $\theta_c$ is the overall phase,
and the spin channel boson field $\theta_v$ is the relative phase
between two components.
While $\eta^\dagger$ depends on the relative phase $\theta_v$,
$O_{qrt}^\dagger$ depends the dual field $\phi_v$.

The overall phase $\theta_c$ is always power-law fluctuating in 1D, and does 
not play a role in the transition. 
It is the relative phase $\theta_v (\phi_v)$ that controls the transition. 
The effective Hamiltonian for this transition is a sine-Gordon theory 
containing cosine terms of both $\theta_v$ and $\phi_v$ as
\bea
H_{res}&=&- \frac{1} {2(\pi a)^2} (\lambda_1 \cos\sqrt{4\pi} 
\phi_v+\lambda_2 \cos\sqrt{4\pi}\theta_v) \ \ \
\mbox{with} \nonumber \\
\lambda_{1,2}&=& (g_t\pm g_v )( \avg{\cos\sqrt{4\pi}\phi_{t1}} 
+\avg{\cos\sqrt{4\pi}\phi_{t2}}).
\eea
In phase $C$ where $\lambda_1> \lambda_2$, then the
relative phase $\phi_v$ is locked giving rise to the pairing order.
In contrast, in phase $B$ where $\lambda_1<\lambda_2$,
the dual field $\theta_v$ is locked giving rise to the quartetting order.
Along the critical phase boundary line 3 in Fig. \ref{fig:RGflow},
$\lambda_1=\lambda_2$ or  $(g_v=0)$, none of $\theta_v$ and $\phi_v$
is pinned.
This transition can be mapped into a theory of two Ising fields
\cite{schulz1996}, one of which is at the critical point.
Equivalently, it can be regarded as a model with two Majorana fermions, 
one of which is massless while the other is massive.
The Ising order and disorder operators are given by
$\cos\sqrt\pi\phi_v$ and $\cos\sqrt\pi\theta_v$ respectively, both of which
have the scaling dimension of $\frac{1}{8}$.

This Ising symmetry breaking effect can also be understood as follows.
In addition to the $SO(5)$ symmetry, spin-3/2 systems
have another $Z_2$ symmetry in the spin channel  $U_n=e^{i n_4 \pi}$
under which $\psi_{\alpha}$, $\eta^\dagger$, and $O_{qrt}$  transform as
\bea
U_n\psi_{\pm \frac{3}{2}} U_n^{-1}&=&i \psi_{\pm \frac{3}{2}}, \ \ \
U_n\psi_{\pm \frac{1}{2}} U_n^{-1}= -i \psi_{\pm \frac{1}{2}}, 
\nonumber \\
U_n \eta^\dagger U_n^{-1}&=&-\eta, \ \ \ \ \ \
U_n O_{qrt} U_n^{-1}= O_{qrt}.
\eea
This $Z_2$ transformation shifts the relative phase $\theta_v$
as $\sqrt \pi \theta_v\rightarrow \sqrt \pi \theta_v \pm \pi$
but leaves $\sqrt \pi \phi_v$ unchanged.
Thus this $Z_2$ symmetry  is broken in the singlet pairing phase B,
but is kept in the quartetting phase C. 

\subsection{Discussion of quartetting in high dimensions}
The study of quartetting phase, or the more general multi-particle
clustering instabilities in high dimensions is a challenging problem.
Ropke {\it et al.} \cite{ropke1998} compared the instabilities of
deuteron (pairing) and $\alpha$-particle (quartetting) channel
in nuclear physics using diagrammatic method.
They found that at low density or strong interaction strength,
the $\alpha$-particle instability wins over the deuteron instability.
Stepnaneko {\it et al.} \cite{stepaneko1999} constructed 
a trial wavefunction in 3D to describe the quartetting
order.
Lee \cite{lee2006} studied the 2D ground state binding energy 
of the $N$-particle cluster in the $SU(N)$ case.

In high dimensions, the competition between quartetting and
pairing superfluidities is more complicated than that in 1D.
First, the pairing superfluidity breaking spin rotational symmetry can not be
stabilized at 1D, but can exist at $D\ge 2$.
Second, phase transitions from weak to strong coupling regimes
can take place at $D\ge 2$ unlike in 1D interactions
always renormalize into strong coupling regime.
Taking into account these facts, we discuss the symmetry class of 
these competitions.
We begin with the $SU(4)$ line with $g_0=g_2=g<0$.
At weak coupling, the sextet pairing state ($\eta^\dagger$ and $\chi^\dagger$
are degenerate) is stabilized breaking the $U(1)$ charge symmetry 
and also the $SU(4)$, or isomorphically $SO(6)$ symmetry in the spin channel.
The residue symmetry is $[SO(5) \ltimes Z_2] \otimes Z_2$.
The  first $Z_2$ is a combined spin-phase operation which flips
the direction of the pairing operator in the spin channel and simultaneously
shift the phase by $\pi$, which is of the same nature responsible for the
HQV in the quintet superfluid in Sect. \ref{sect:quintet}.
The second $Z_2$ is  purely the phase operation  
$\psi_\alpha\rightarrow -\psi_\alpha$ which leave pairing operators unchanged.
On the other hand, the quartetting state in the strong coupling limit
is $SU(4)$ invariant, and also breaks the $U(1)$ symmetry.
Its residue symmetry is $SO(6)\otimes Z_4$ where $Z_4$
operation means $\psi_\alpha\rightarrow e^{i m \pi/2} \psi_\alpha
(m=1\sim 4)$.
Thus this phase transition is describe by the coset
$SO(6)\otimes Z_4/ [SO(5) \ltimes Z_2 \otimes Z_2]= S^5$.
In other words, the pairing to quartetting transition can be viewed
as an order to disorder transition of pairing operators for their
spin configuration on the $S^5$ sphere.
If the spin configuration becomes disordered, due to the combined 
spin-phase $Z_2$ structure, the phase of the pair operators 
is only well defined modulo of $\pi$, thus it means the quartetting
order appears.
If $g_2<g_0< 0$ or $g_0<g_2<0$, then in the weak coupling limit, the quintet
pairing state $\chi^\dagger$, or the singlet pairing $\eta^\dagger$
dominates.
A similar reasoning shows that the coset for the
transition from the quartetting state to the quintet or singlet
Cooper pairing states is $S^4$ or $Z_2$, respectively.

%******************************************************************
\section{Magnetic properties in spin-3/2 systems}
\label{sect:magnetism}
In the strong repulsive interaction regime $U_0,~U_2\gg t$
and at commensurate fillings, the low energy physics of
spin-3/2 systems are described by the
magnetic exchange models.
In this section, we review the magnetic properties in such
systems\cite{wu2005a,chen2005}, and also present new results
in Sect. \ref{subsect:2dquarter} and \ref{subsect:half}.

\subsection{Magnetic exchange at quarter-filling}
We first construct the exchange Hamiltonian at quarter-filling
i.e., one particle per site. 
The total spin of two sites in a bond can be $S_{tot}=0,1,2,3$.
Using the projection perturbation theory, we find the exchange
exchange energy in each channel as
\bea
J_0= \frac{4t^2}{U_0}, \ \ \ J_2= \frac{4t^2}{U_2}, \ \ \ J_1=J_3=0,
\eea
where the degeneracy of $J_{1,3}$ is a consequence of the $SO(5)$
symmetry.
This model can be represented in the standard Heisenberg
type by using bi-linear, bi-quadratic, and bi-cubic terms as
$
H_{ex}=\sum_{ij} a~ (\vec S_i \cdot \vec S_j)+
b~ (\vec S_i \cdot \vec S_j)^2+ c~ (\vec S_i \cdot \vec S_j)^3,
$
with $a=-\frac{1}{96}(31 J_0+23 J_2)$, $b=\frac{1}{72}(5 J_0+17 J_2)$
and $c=\frac{1}{18}(J_0+ J_2)$.
More elegantly, it can be represented in the explicit $Sp(4)$
symmetric form as
\bea
H_{ex}&=& \sum_{ij}\Big\{ \frac{J_0+J_2}{4} L_{ab} (i) L_{ab} (j) 
+ \frac{3J_2-J_0}{4}  n_a(i) n_a(j) \Big\}.
\label{eq:exchange}
\eea

Eq. \ref{eq:exchange} satisfies two different $SU(4)$ symmetries 
at $J_0=J_2$ and $J_2=0$, respectively.
The first one denoted as $SU(4)_A$ below
is obvious at $J_0=J_2$ where Eq. \ref{eq:exchange} reduces into 
the $SU(4)$ Heisenberg model of
$H_{SU(4),A}
=\sum_{ij} \frac{J}{2}\Big\{ L_{ab} (i) L_{ab} (j) 
+   n_a(i) n_a(j) \Big\}.$
Each site is in the fundamental representation.
The second $SU(4)$ symmetry exists in the bipartite
lattice at $J_2=0$ denoted as $SU(4)_B$ below.
We perform a particle-hole transformation in the odd sublattice
$L^\prime_{ab}= L_{ab}, ~ n^\prime_a=-n_a$,
i.e., the odd sites are transformed into the 
anti-fundamental representation.
Eq. \ref{eq:exchange} is again $SU(4)$ invariant as
\bea
H_{SU(4),B}&=& \sum_{ij} \frac{J_0}{4} \Big\{  L^\prime_{ab} (i) L_{ab} (j) 
+ n^\prime_a(i) n_a(j) \Big\},
\eea
but with fundamental and anti-fundamental representations alternatively
in even and odd sites.
This $SU(4)_B$ model and the more generally staggered $SU(N)$ Heisenberg
model were investigated extensively by using the large $N$ method
\cite{arovas1988,read1990}.
These two $SU(4)$ symmetries have dramatically different physical properties.
At least four sites are needed to form an $SU(4)$ singlet for the $SU(4)_A$,
while for $SU(4)_B$, two sites are enough to form a bond singlet.

\subsection{1D phases  at quarter-filling}
The phase diagram of Eq. \ref{eq:exchange} at 1D can be obtained
by using the  bosonization method to the Hubbard model.
At quarter-filling, the $8k_f$ term Umklapp term is important and
only involves the charge sector.
It can be bosonized as
\bea
{\cal H}_{um,8k_f}&=& \frac{g_u}{2(\pi a)^2} \cos (\sqrt {16 \pi}
\phi_{c}-8 k_f x ),
\label{eq:um8kf}
\eea 
where $g_u$ is at the order of $O(g^3_0,g^3_2)$ at the bare level.
At quarter-filling, the spin channels behave the same as in
the incommensurate case discussed in Sect. \ref{subsect:1dphase}.
The charge channel opens up the gap at $K_c<1/2$
where $g_u$ is renormalized to infinity.
As a result, the degeneracy between the
real and imaginary parts of the $O_{2k_f,cdw}$ and $O_{4k_f,cdw}$,
are lifted.
Their real parts describe the usual CDW orders, while their 
imaginary parts mean the bond orders, i.e., the $2k_f$ and $4k_f$ 
spin Peierls orders.

With the opening of charge gap at quarter-filling, various insulating
phases appear as given in Fig. \ref{fig:1dphase}.
Phase A and C1 corresponds to the regime described by the
exchange model Eq. \ref{eq:exchange} with $J_2\ge J_0$ and $J_2<J_0$,
respectively.
Phase A ($J_2\ge J_0>0$), including the $SU(4)_A$ line,
is a gapless spin liquid phase.
The $SU(4)_A$ line was solved by Sutherland \cite{sutherland1975}.
In phase C.1 ($J_2\le J_0$), the spin gap opens with the
developing of the $4k_f$ spin-Peierls order, i. e., the dimer order.
The transition between these two phases is
Kosterlitz-Thouless like.
Insulating phases involving CDW order also exist
in phases B.1, B.2 and C.2.
The quartetting phase B splits into two parts B.1 and B.2. 
In phase B.1, the $SU(4)$ singlet quartets exhibit both the charge 
and spin Peierls orders.
In phase B.2, the 2$k_f$ CDW of quartets becomes long range ordered.
Similarly, the CDW of singlet pairs becomes long range ordered
in phase C.2. 
The boundaries between phase B.1 and B.2, phase C.1 and C.2 are
determined by the bare value of $g_{u0}=0$ as sketched
in Fig. \ref{fig:1dphase}.
However, due to the non-universal relations between $g_{u0}$ and
$g_{0,2}$, the exact boundaries are hard to determine.

\subsection{Discussion of the 2D phase diagram at quarter-filling}
\label{subsect:2dquarter}

The 2D physics of Eq. \ref{eq:exchange} is a challenging problem.
Nevertheless, good understanding has been achieved for the $SU(4)_B$ line
($J_2=0$) in the square lattice where both quantum Monte-Carlo 
simulations \cite{harada2003} and the large-$N$ analysis show
that the ground state \cite{read1990,zhang2001}
is long range Neel ordered.
As a result of strong quantum fluctuations, the Neel moments are 
$(-)^i n_4 =(-)^i L_{15} =(-)^i L_{34}\approx 0.05$,
which is really tiny compared to the $SU(2)$ case.
The Goldstone manifold is $CP(3)=U(4)/[U(1) \otimes U(3)]$
with 6 branches of spin-waves.

\begin{figure}
\centerline{\psfig{file=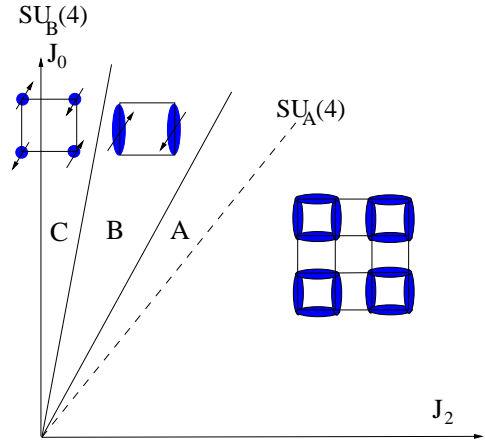,width=0.5\linewidth,angle=0}}
\caption{Speculated phase diagram at quarter-filling in 
a 2D square lattice.
Phase A is the plaquette ordered phase which is suggested
along the $SU(4)_A$ line in an exact diagonalization 
work by Bossche et al. $^{67}$
%\cite{bossche2000}.
Phase C is the Neel ordered phase which is confirmed along
the $SU(4)_B$ line in QMC simulations by Harada et al.
$^{65}$
%\cite{harada2003}.
We speculate a dimerized phase B lying between phase A and C.
Dimers in phase B carry antiferromagnetic spin-nematic order.
}
\label{fig:2Ddiagram}
\end{figure}

However, the physics is less clear away from the $SU(4)_B$ line.
The $SU(4)_A$ line of $J_0=J_2$ is of particular interest.
In order to form a singlet for $SU(4)_A$, we need four sites
around a plaquette with the wavefunction of
 $\frac{1}{4!}\epsilon_{\alpha\beta\gamma\delta} 
\psi^\dagger_\alpha(1) \psi^\dagger_\beta(2)  \psi^\dagger_\gamma(3) 
\psi^\dagger_\delta(4) |\Omega\rangle$ where $|\Omega\rangle$
is the vacuum.
Indeed, an exact diagonalization by Bossche {\it et al.}
in a $4\times 4$ sites lattice \cite{bossche2000} suggests 
that the ground state along the $SU(4)_A$ line exhibits
such plaquette order
as  depicted in Fig. \ref{fig:2Ddiagram}.
This plaquette order was also found in a large-$N$ mean field
theory by Mishra {\it et al} \cite{Mishra2002}.
However, due to the sign problem in QMC,
a large size simulation is difficult to confirm this result.
Recently, Chen {\it et al} \cite{chen2005} constructed an
$SU(4)$ Majumdar-Ghosh model in a two-leg spin-3/2 
ladder whose ground state is solvable exhibiting
this plaquette state.
Similarly to the quartetting order in the superfluid state, 
this plaquette order is of four sites without any site and bond spin orders.
Upon doping, we suggest that a quartetting superfluid 
appears.

Now let us speculate the phase diagram in the entire parameter
regime as depicted in  Fig. \ref{fig:2Ddiagram}.
We expect that the long range Neel order  can 
survive with a finite value of $J_2$ as marked as phase 
C in Fig. \ref{fig:2Ddiagram}.
The plaquette ordered phase is gapped, thus it should be stabilized
in an entire phase as marked phase C.
It extends into a finite regime with $J_2<J_0$, and we also
speculate that it survives in the entire region of $J_0<J_2$.
Furthermore, between phase C with the site spin order  and
phase A with the plaquette spin order, we  suggest the existence of 
the magnetically ordered dimer phase B.
The dimer made of two spin-3/2 particles
possesses the internal spin structures.
In phase B where $J_2<J_0$, the energy of the dimer singlet state
is lower than the energy of the dimer quintet.
Their superposition might give rise to an antiferromagnetically
ordered spin-nematic dimer state.
More analytic and numeric works are desired to confirm 
these speculations. 

\subsection{spin-3/2 magnetism at half-filling}
\label{subsect:half}
\begin{figure}
\centering\epsfig{file=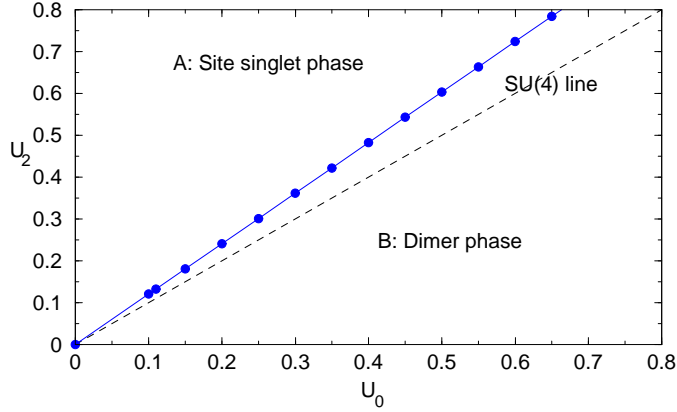,clip=1,width=0.7\linewidth,angle=0}
\caption{1D phase diagram at half-filling (two particles per site)
with $U_{0,2}>0$ ($t$ is rescaled to 1).
The phase boundary is marked with the solid line,
and the $SU(4)$ line ($U_0=U_2$) is marked with the dashed line.
A) The site singlet phase; B) The spin-Peierls (dimer) phase.
}\label{fig:1dhalf}
\end{figure}

At half-filling with $U_0, U_2\gg t$, the exchange model
has been derived as  Eq. \ref{eq:heisenberg}.
In 1D, the phase diagram of Eq. \ref{eq:heisenberg} can
be obtained by applying the  bosonization method directly to 
the Hubbard model after taking into account the  $4k_f$-Umklapp terms
as
\bea 
{\cal H}_{um, 4k_f}&=& \frac{\lambda_s}{2} \eta^\dagger_R
\eta_L +\frac{\lambda_v}{2} \chi^a_R \chi^a_L +h.c. ,
\eea
where $\eta_R=\psi^\dagger_{R\alpha} R_{\alpha\beta} \psi^\dagger_{R\beta}$,
$\chi_R=\psi^\dagger_{R\alpha} (R\Gamma^a)_{\alpha\beta} 
\psi^\dagger_{R\beta}$, and $\eta_L,\chi_L$ are defined correspondingly.
The bare values for $\lambda_{s,v}$  are 
$\lambda_s=U_0 a_0,  \lambda_v=U_2 a_0$ where $a_0$ is the
lattice constant.
This term can be bosonized as
\bea\label{um4kf} {\cal
H}_{um,4k_f}&=&- \frac{1}{2(\pi a)^2} \cos (\sqrt {4\pi} \phi_{c})
\nonumber  \Big\{ (\lambda_v+\lambda_s)
\cos\sqrt{4\pi}\phi_v +(\lambda_v-\lambda_s) \cos\sqrt{4\pi}\theta_v
\nonumber \\ &+& 2\lambda_v ( \cos\sqrt {4\pi} \phi_{t1} +\cos\sqrt
{4\pi} \phi_{t2}) \Big\}, \eea 
Together with Eq. \ref{eq:1dHam}, the RG equations at the half-filling
can be calculated as
\bea
\frac{ d g_c}{d \ln t}&=& \frac{1}{2\pi}
(\lambda_s^2+ 5\lambda_v^2), \hspace{2cm}
\frac{ d g_v} {d \ln t} =
\frac{1}{2\pi} ( 4 g_v g_t + 2\lambda_s \lambda_v), \nonumber \\
\frac{ d g_t}{d \ln t}&=& \frac{1}{2\pi} ( 3 g_t^2 + g_v^2 + 2
\lambda_v^2),   \hspace{1.1cm}
\frac{ d \lambda_s} {d \ln t}=
\frac{1}{2\pi} (g_c\lambda_s + 5 g_v \lambda_v), \nonumber \\
\frac{ d \lambda_v} {d \ln t}&=&
\frac{1}{2\pi} (g_c \lambda_v + g_v \lambda_s
+4 g_t \lambda_v). 
\label{eq:rghalf} 
\eea
Eq. \ref{eq:rghalf} are solved numerically in the region $U_0>0, U_2>0$.
Two stable fixed points are found as 
\bea
\mbox{A}: g_c=-g_v=g_t=-\lambda_s=\lambda_v\rightarrow +\infty, \ \ \
\mbox{B}: g_c=g_v=g_t=\lambda_s=\lambda_v\rightarrow +\infty. \ \ \
\eea
with the corresponding phase diagram shown in Fig. \ref{fig:1dhalf}.
Phase A is the site singlet phase with the pinned values of the boson 
fields as $\avg{\phi_c}=\avg{\theta_v}=\avg{\phi_{t1}}=\avg{\phi_{t2}}=0$.
All the charge orders and spin-Peierls orders vanish in this phase.
Phase B is the dimerized phase with pinned boson fields as
$\avg{\phi_c}=\avg{\phi_v}=\avg{\phi_{t1}}=\avg{\phi_{t2}}=0$.
As a result, the 2$k_f$ spin-Peierls (dimer) order is long range
ordered.
We notice that the $SU(4)$ line of $U_0=U_2$ is 
already inside the dimer phase, thus is not critical.
This agrees with the previous RG result in Ref. \cite{assaraf2004}.
The phase boundary is determined by numerically solving 
the above RG equations, which extends to the region
with $U_0<U_2$.
The transition between phase A and B is Ising-like,
which is driven by the competitions between $\phi_v$ and $\theta_v$.

The two dimension phase diagram 
of Eq. \ref{eq:heisenberg}  is also challenging.
As discussed in Sect. \ref{subsect:strong}, at $U_0<U_2$
Eq. \ref{eq:heisenberg} reduces to the $SO(5)$ rotor model.
When the difference between $U_2$ and $U_0$ is small,
or large compared to the inter-site exchange,
the system is in the antiferromagnetic spin-nematic phase,
or in the site singlet phase.
In the regime of $U_2 \le U_0$, it is not clear whether the ground state
still possesses magnetic long range order.
An interesting numeric QMC result \cite{assaad2005} showed that 
along the $SU(4)$ line, Eq. \ref{eq:heisenberg} is in
the gapless spin liquid phase, but more numeric work is desirable
to confirm it.

%*****************************************************************
\section{Summary}
\label{sect:summary}
In summary, we have reviewed the hidden $SO(5)$ symmetry in 
spin-3/2 cold atomic systems.
This symmetry is proved to be exact in the continuum model with
$s$-wave interactions, and in the lattice Hubbard model in the
optical lattices.
This high symmetry provides a framework to understand
various properties in spin-3/2 systems, including
the Fermi liquid theory,  mean field phase diagram, 
and the sign problem in QMC simulations.
spin-3/2 systems support novel superfluid states including
the quintet Cooper pairing state and the quartetting state.
The topological defect of half-quantum vortex and the $SO(4)$
Cheshire charge effect were discussed  in the quintet superfluid.
The existence of the quartetting state in 1D systems and its
competition with the Cooper pairing state were investigated.
At last, we reviewed the magnetism in spin-3/2 systems
showing many different features from the spin-$\frac{1}{2}$ systems.

Taking into account the rapid progress in the cold atomic
physics, we are optimistic that the spin-3/2 high spin
cold atomic systems can be realized in experiment in the near future.
We hope that the research on the hidden symmetry aspect
can stimulate general interest in various properties 
in such systems.

% Previously published material must be accompanied by written 
% permission from the author and publisher.
\section*{Acknowledgments}
I thank my Ph. D advisor S. C. Zhang for his guidance on
the research in spin-3/2 cold atoms,
and L. Balents, C. Capponi, S. Chen, J. P. Hu,
O. Motrunich, Y. P. Wang for collaborations on this topic.
I  also thank H. D. Chen, E. Demler, L. M. Duan, E. Fradkin, 
M. P. A. Fisher, T. L. Ho, A. J. Leggett, M. Ma, D. Scalapino, M. Troyer, 
J. Zaanen, and F. Zhou for helpful discussions.
This work is supported by the NSF Grant No. Phy99-07949.

%\bibliographystyle{prsty}
%\bibliography{spin32,extra}

\end{document}